\ifdefined\arxivpreprint
  \documentclass[aps,prl,preprint,superscriptaddress,floatfix,nolongbibliography]{revtex4-2}
\else
  \documentclass[aps,prl,reprint,superscriptaddress,floatfix,nolongbibliography]{revtex4-2}
\fi

\usepackage{graphicx}
\usepackage{dcolumn}
\usepackage{bm}
\usepackage{amsmath,amssymb}
\usepackage[colorlinks=true,citecolor=blue,linkcolor=blue,urlcolor=blue]{hyperref}

\newcommand{\bra}[1]{\langle #1 |}  
\newcommand{\ket}[1]{| #1 \rangle}  
\begin{document}


\title{State-Dependent Diffusion and Spectra of Strongly Driven Thermal Atoms}


\author{Zheng Xiao}
\thanks{These authors contributed equally to this work.}
\affiliation{School of Electronics, Peking University, Beijing 100871, China}
\author{Zijie Liu}
\thanks{These authors contributed equally to this work.}
\affiliation{School of Electronics, Peking University, Beijing 100871, China}
\author{Suyang Wei}
\affiliation{School of Electronics, Peking University, Beijing 100871, China}
\author{Anhong Dang}
\email{ahdang@pku.edu.cn}
\affiliation{School of Electronics, Peking University, Beijing 100871, China}
\author{Tiantian Shi}
\email{tts@pku.edu.cn}
\affiliation{National Key Laboratory of Advanced Micro and Nano Manufacture Technology, School of Integrated Circuits, Peking University, Beijing 100871, China}
\affiliation{Beijing Advanced Innovation Center for Integrated Circuits, Beijing 100871, China}
\author{Jingbiao Chen}
\affiliation{School of Electronics, Peking University, Beijing 100871, China}
\affiliation{National Key Laboratory of Advanced Micro and Nano Manufacture Technology, School of Integrated Circuits, Peking University, Beijing 100871, China}
\affiliation{Hefei National Laboratory, Hefei 230088, China}

\date{\today}

\begin{abstract}
We propose a state-dependent diffusion model for strongly driven thermal-atom spectra. Starting from the trajectory-dependent internal-state evolution of individual atoms, we derive a closed spatial equation for the local density-matrix field using a velocity-moment expansion. Measurements of an $^{85}$Rb atomic-filter transmission spectrum agree well with the model up to a maximum Gaussian peak intensity of $1.27\times10^{3}$ W/cm$^2$, approaching six orders of magnitude above the $^{85}$Rb D2-line saturation intensity. Counterintuitively, the model reveals an anomalous optical-pumping pathway in which intense light transfers atoms from nominally dark states into bright states. Hyperfine Paschen--Back splitting selectively enhances this anomalous pathway while suppressing conventional optical pumping, allowing the filter to maintain approximately 97$\%$ transmission at the highest intensity studied. This work provides a framework for controlling strongly driven atomic ensembles and designing saturation-resistant atomic optical devices.
\end{abstract}


\maketitle

A fundamental challenge in nonequilibrium physics is to derive experimentally accessible continuum fields from the microscopic dynamics of many particles \cite{hilbert1902mathematical,deng2025hilbert}. In light--atom interactions, this challenge has a quantum counterpart: the collective internal state and optical response of a thermal ensemble must be derived from the internal-state dynamics of moving atoms \cite{cohentannoudji1992atom,auzinsh2010optically}. Solving this problem is essential for accurately describing the absorptive and dispersive properties of atomic media. These narrowband responses, which are set by the atomic level structure, underpin atomic clocks and chip-scale frequency standards \cite{ludlow2015optical,takamoto2005optical,knappe2004microfabricated,kitching2018chip}, atomic magnetometers, high-power alkali-vapor lasers, and frequency-stabilized laser systems \cite{budker2007optical,kominis2003subfemtotesla,wieman1976doppler,yang2011power,krupke2012dpal}, slow-light and quantum-memory devices \cite{hau1999light,kash1999ultraslow,fleischhauer2000dark,julsgaard2004experimental}, and narrowband atomic filters for free-space communication and strong-background rejection \cite{menders1991ultranarrow,tang1995experimental,frickebegemann2002daylight,yin2024ghost}.

Unlike a classical number density, the local state of a thermal atomic ensemble is a density matrix whose populations and coherences are jointly determined by coherent driving, spontaneous emission, collisions, and transport across an inhomogeneous optical field \cite{happer1972optical,berman1982collision,firstenberg2013coherent}. Existing descriptions of this coupled problem fall into two main categories. Time-domain master-equation and rate-equation methods represent atomic motion through finite interaction times; by retaining multilevel internal dynamics, they have successfully described saturation and nonlinear optical pumping under strong driving \cite{maguire2006theoretical,lindvall2009interaction,vanlange2020combined,haupl2025modelling}, but do not yield a closed spatial evolution equation for transport across an inhomogeneous beam. Spatial models instead represent atomic transport by adding a scalar diffusion term, $D\nabla^2\rho$, to the local density-matrix equation \cite{torrey1956bloch,firstenberg2008theory}. They have elucidated Dicke narrowing \cite{dicke1953collisions,firstenberg2007dicke}, transit-time broadening in finite beams \cite{borde1976transit,chekhovskoi2024transit}, diffusion-induced Ramsey narrowing \cite{xiao2006diffusion,tilchin2011thermal}, and diffusion of stored optical coherence \cite{shuker2008storing,firstenberg2010selfsimilar,finkelstein2023practical}, but prescribe the same transport law for all internal-state modes. Consequently, neither category provides a general description of the differential transport generated by the coupled effects of strong optical driving, relaxation, and atomic motion.

In this work, we start from a semiclassical phase-space master equation and use a velocity-moment expansion to derive a closed equation for the local density-matrix field. In this equation, atomic transport is intrinsically coupled to the local multilevel dynamics generated by coherent optical driving, spontaneous emission, and collisional relaxation; the resulting transport operator is consequently mode dependent rather than a prescribed scalar diffusion coefficient. We validate the model against measured transmission spectra of a $^{85}$Rb atomic filter at intensities approaching six orders of magnitude above the D2-line saturation intensity. The calculated populations further reveal an anomalous optical-pumping mechanism: intense light transfers atoms from nominally dark states into bright states and thereby suppresses saturation of the principal transition. Hyperfine Paschen--Back splitting selectively enhances anomalous optical pumping while suppressing conventional optical pumping, allowing the filter to maintain approximately 97$\%$ transmission at the highest intensity studied.

\begin{figure}
\centering
\ifdefined\arxivpreprint
  \includegraphics[width=0.62\linewidth]{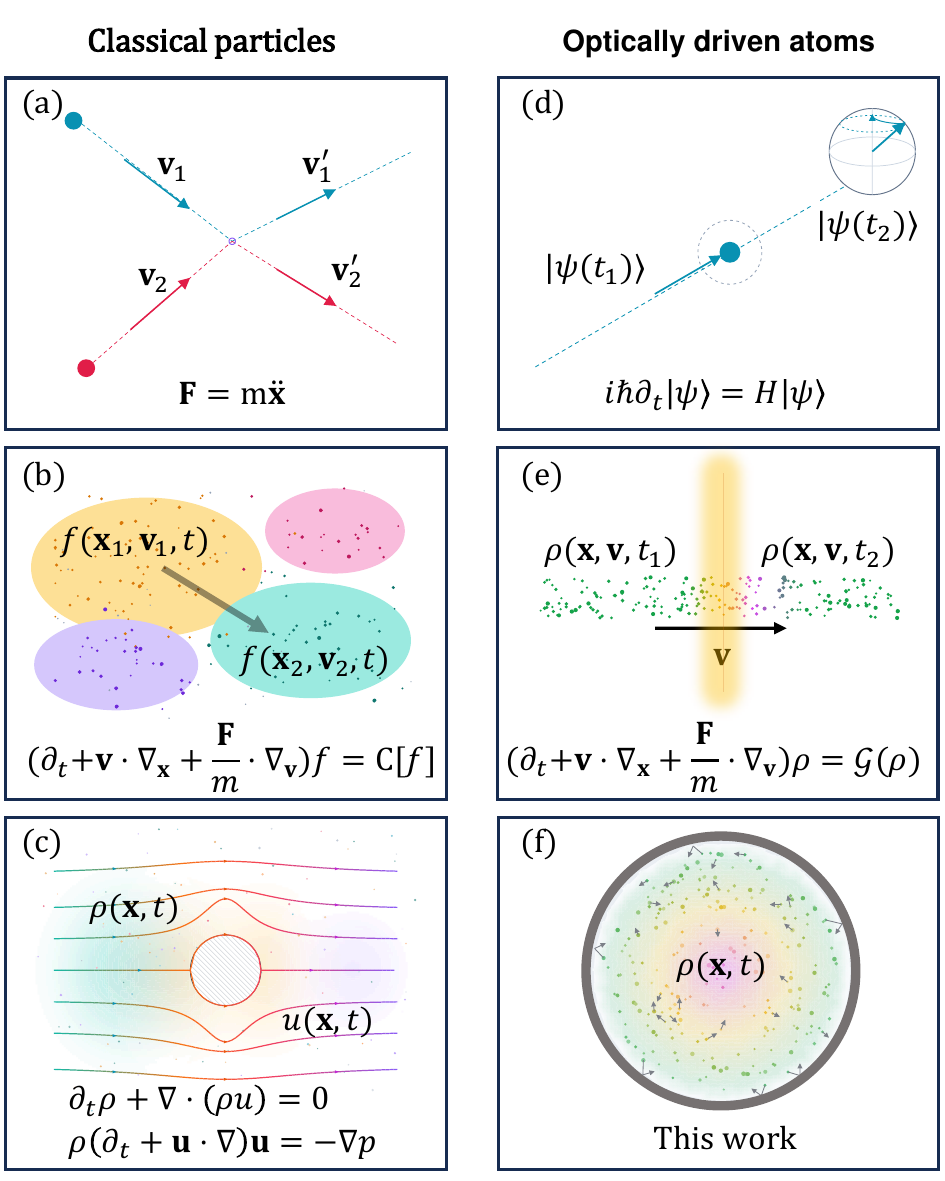}
\else
  \includegraphics[width=0.82\linewidth]{fig1.pdf}
\fi
\caption{\label{fig:fig1} Hierarchy of descriptions for classical particles (left column) and atoms interacting with light (right column). (a)--(c) Classical hierarchy: deterministic single-particle trajectories (a, Newtonian mechanics) $\rightarrow$ a phase-space distribution (b, Boltzmann equation) $\rightarrow$ macroscopic continuum fields (c, fluid dynamics). (d)--(f) Semiclassical atomic hierarchy: single-atom internal-state evolution (d, Schr\"odinger equation) $\rightarrow$ a trajectory-resolved density-matrix distribution (e, phase-space master equation) $\rightarrow$ a local density-matrix field (f, diffusion-form master equation). Both columns proceed from a microscopic description through a kinetic description to a macroscopic field equation. This work derives the reduction (e)$\rightarrow$(f).}
\end{figure}

We derive the spatial evolution of the local density-matrix field from the trajectory-dependent internal-state dynamics of individual atoms, as illustrated by the reduction from Fig.~\ref{fig:fig1}(e) to Fig.~\ref{fig:fig1}(f). Let $\varrho(t,\mathbf{x},\mathbf{v})$ denote the density-matrix-valued distribution at the phase-space point $(\mathbf{x},\mathbf{v})$. We consider motion in the two-dimensional plane transverse to the optical axis. The atomic center-of-mass motion is treated semiclassically, whereas the internal state evolves under the local generator $\mathcal{G}$, which includes coherent driving, spontaneous emission, and collisional relaxation:
\begin{equation}
\frac{\partial \varrho}{\partial t}+\mathbf{v}\cdot\nabla_{\mathbf{x}}\varrho+\frac{\mathbf{F}}{m}\cdot\nabla_{\mathbf{v}}\varrho=\mathcal{G}(\varrho).
\label{eq:1}
\end{equation}
Here $\mathbf{x}$, $\mathbf{v}$, and $\mathbf{F}$ are the transverse position, velocity, and optical force, respectively. To obtain the local ensemble state, we expand the phase-space distribution in velocity moments. When the characteristic transverse scale $R$ of the optical field is much larger than the relaxation length $v_{\mathrm{th}}/\Gamma_{\mathrm{rel}}$, collisions maintain a local Maxwell distribution and higher velocity moments relax rapidly. Retaining the isotropic and first angular moments gives
\begin{equation}
\varrho=\frac{m}{2\pi k_B T}\,e^{-mv^{2}/(2k_BT)}
\left[\bar{\varrho}+\frac{m\mathbf{v}}{k_BT}\cdot\mathbf{J}\right],
\label{eq:2}
\end{equation}
where $v_{\mathrm{th}}=\sqrt{2k_BT/m}$, $\bar{\varrho}=\int\!\varrho\,d^{2}v$ is the local density-matrix field (the zeroth moment), and $\mathbf{J}=\int\!\mathbf{v}\varrho\,d^{2}v$ is its flux (the first moment). In particular, $\operatorname{Tr}\bar{\varrho}=n$ is the local number density; thus $\bar{\varrho}$ is not normalized to unit trace. Taking the zeroth and first moments of Eq.~(\ref{eq:1}) yields
\begin{eqnarray}
\frac{\partial \bar{\varrho}}{\partial t}+\nabla\cdot\mathbf{J} &=& \mathcal{G}(\bar{\varrho}),\label{eq:3}\\
\frac{\partial \mathbf{J}}{\partial t}+\nabla\left(\frac{k_BT}{m}\bar{\varrho}\right)-\frac{\mathbf{F}}{m}\bar{\varrho} &=& \mathcal{G}(\mathbf{J}).\label{eq:4}
\end{eqnarray}
Equation~(\ref{eq:3}) is the zeroth-moment balance equation: the local internal state changes through both local dynamics and the divergence of the atomic flux. Equation~(\ref{eq:4}) is the corresponding flux equation; gradients of density and temperature and the optical force drive the flux, whereas relaxation is represented by $\mathcal{G}(\mathbf{J})$. When the flux relaxes much faster than $\bar{\varrho}$ evolves, $\mathbf{J}$ can be adiabatically eliminated. Because particle-number conservation gives $\mathcal{G}$ a zero mode, we define the reduced inverse $\mathcal{G}_{Q}^{-1}$ on the complementary subspace $Q$ of relaxing modes. The diffusion-form master equation then becomes
\begin{equation}
\frac{\partial \bar{\varrho}}{\partial t}+\nabla\cdot\left\{\mathcal{G}_{Q}^{-1}
\left[\nabla\left(\frac{k_BT}{m}\bar{\varrho}\right)-\frac{\mathbf{F}}{m}\bar{\varrho}\right]\right\}
=\mathcal{G}(\bar{\varrho}).
\label{eq:5}
\end{equation}
The operator $\mathcal{G}_{Q}^{-1}$ allows different population and coherence modes to acquire different transport length scales. The fields $n$ and $T$ are determined at the macroscopic level: the optical force redistributes the atoms, while absorption heats the gas and causes $T$ to depart from its boundary value. In steady state, they obey the force-balance and heat equations
\begin{eqnarray}
\nabla(k_BT\,n) &=& \mathbf{F}\,n,\label{eq:6}\\
\nabla\cdot(\kappa\nabla T)+Q &=& 0.\label{eq:7}
\end{eqnarray}
where $\kappa$ is the thermal conductivity and $Q$ is the absorbed optical power per unit volume. In the isothermal limit $T\equiv T_0$---appropriate when optical heating is weak or wall-mediated thermalization is rapid---Eq.~(\ref{eq:6}) gives the Boltzmann distribution $n=n_0e^{-U/(k_BT_0)}$ for $\mathbf{F}=-\nabla U$. If all relevant flux modes share a common transport-relaxation rate $\Gamma_{\mathrm{tr}}$, or if the linear response is dominated by a single internal-state mode, then $\mathcal{G}_{Q}^{-1}\rightarrow-\Gamma_{\mathrm{tr}}^{-1}$ is effectively scalar and Eq.~(\ref{eq:5}) reduces, in the isothermal force-free limit, to
\begin{equation}
\frac{\partial \bar{\varrho}}{\partial t}=\mathcal{G}(\bar{\varrho})
+\frac{k_BT_0}{m\Gamma_{\mathrm{tr}}}\nabla^{2}\bar{\varrho}.
\label{eq:8}
\end{equation}
Thus all relevant internal-state modes share the scalar diffusion coefficient $D=k_BT_0/(m\Gamma_{\mathrm{tr}})$. In a strongly driven multilevel system, different population and coherence modes generally have different relaxation eigenvalues, and $\mathcal{G}_{Q}^{-1}$ cannot be replaced by a prescribed scalar coefficient. Once $\bar{\varrho}$ has been obtained, the positive-frequency polarization and the effective susceptibility tensor are defined by
\begin{equation}
\mathcal{P}^{(+)}_q=\sum_{ij}\bra{g_j}d_q\ket{e_i}
\bar{\varrho}_{e_i g_j}
=\varepsilon_0\sum_{q'}\chi_{qq'}\mathcal{E}_{q'}.
\label{eq:9}
\end{equation}
Here $d_q$ and $\mathcal{E}_q$ are spherical components of the electric-dipole operator and the complex optical-field amplitude, respectively, and $\bar{\varrho}_{e_i g_j}$ is an optical coherence. For an isolated transition, Eq.~(\ref{eq:9}) reduces to the familiar response proportional to
$|d_{ij}^{(q)}|^2(\bar{\varrho}_{g_jg_j}-\bar{\varrho}_{e_ie_i})/(\Delta_{ij}+i\gamma_{ij})$, where $\Delta_{ij}=\omega-\omega_{ij}$. Axial motion produces Doppler shifts; we calculate the susceptibility for each axial-velocity class and then average over the axial Maxwell distribution. The full derivation and the definitions of the internal-state parameters are given in the Supplemental Material. We now examine steady-state solutions of Eq.~(\ref{eq:5}) for representative multilevel systems.

\begin{figure}
\includegraphics[width=1\linewidth]{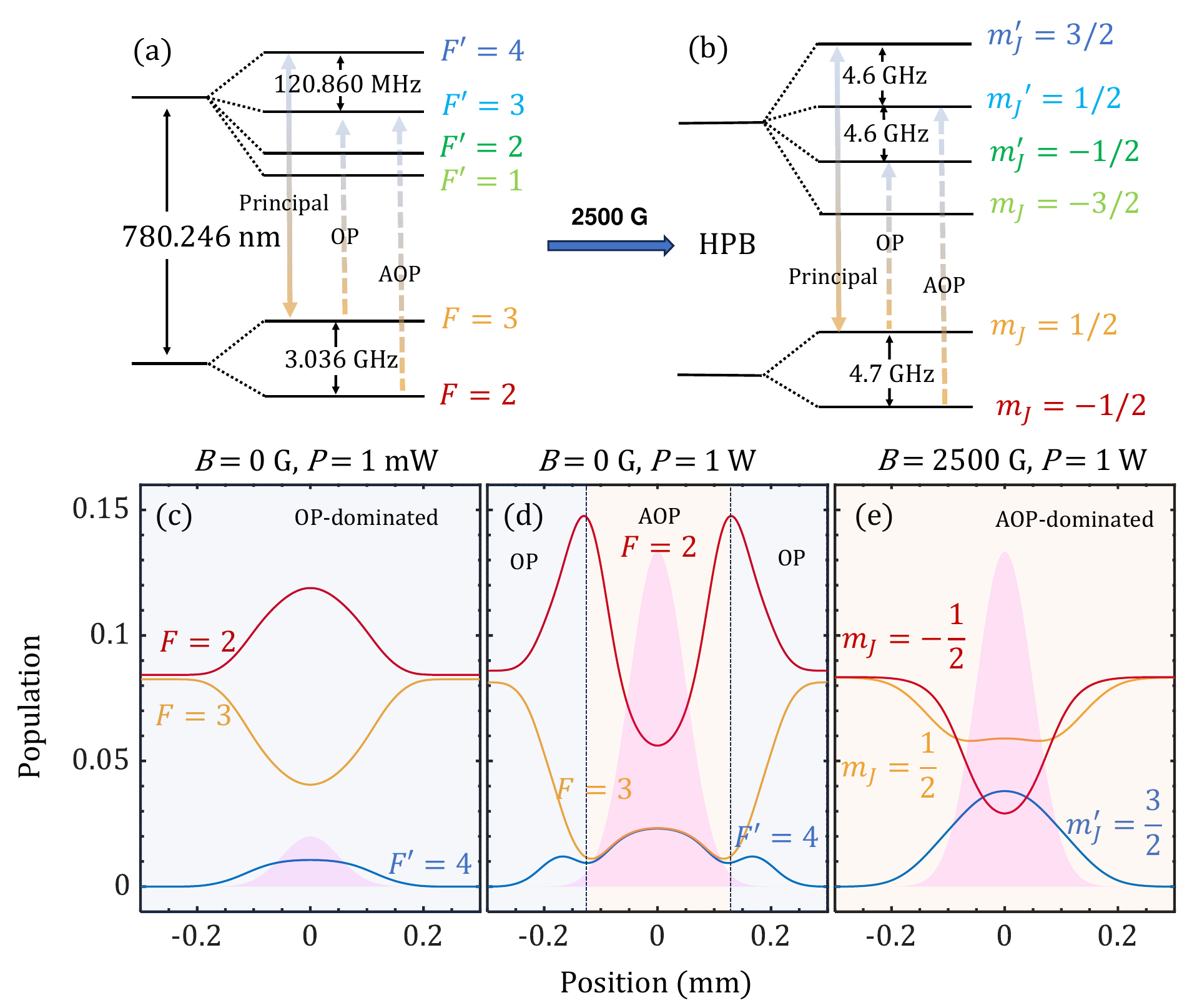}
\caption{\label{fig:fig2} Spatially resolved population dynamics of $^{85}$Rb driven near the 780.246-nm cycling transition. (a) Relevant zero-field hyperfine levels for the $F=3\rightarrow F'=4$ transition. (b) Level structure in a 2500-G axial magnetic field, where the atom is in the hyperfine Paschen--Back regime and the states are grouped by $m_J$ and $m_J'$. (c)--(e) Calculated position-dependent populations in a Gaussian beam: (c) 1 mW at zero magnetic field, (d) 1 W at zero magnetic field, and (e) 1 W in the hyperfine Paschen--Back regime. In the annotations, ``Principal'' identifies the principal transition; ``OP'' and ``AOP'' identify transitions or spatial regions dominated by conventional and anomalous optical pumping, respectively. At high intensity in the hyperfine Paschen--Back regime, anomalous optical pumping transfers population from nominally dark states into the bright cycling-state manifold.}
\end{figure}

We call the transition group closest to resonance and dominant in the optical response the principal transition. Ground states coupled by this transition are bright states; the remaining far-off-resonant ground states are nominally dark states. Here, ``dark' denotes far-off-resonant ground states, not coherent dark-state superpositions. We first consider $^{85}$Rb at zero magnetic field, tune a single-mode laser to the $F=3\rightarrow F'=4$ cycling transition [Fig.~\ref{fig:fig2}(a)], and calculate the spatial population distributions at different optical powers.

As shown in Fig.~\ref{fig:fig2}(c), at a laser power of 1 mW the spatial populations exhibit conventional optical pumping. The $F'=4$ level is separated from the neighboring $F'=3$ level by only 120.860 MHz \cite{steck2025rb85}, so Doppler and power broadening allow off-resonant excitation on $F=3\rightarrow F'=3$. Subsequent spontaneous decay transfers population from the bright $F=3$ manifold to the nominally dark $F=2$ manifold. Consequently, the $F=3$ population is depleted at the beam center, whereas the $F=2$ population increases. When the laser power is raised to 1 W, the population distribution changes qualitatively. Conventional bright-to-dark optical pumping persists in the low-intensity region near the beam edge [light-blue region in Fig.~\ref{fig:fig2}(d)]. In the high-intensity central region (light-yellow region), however, the principal $F=3\rightarrow F'=4$ transition is saturated, and the combined population of its lower and upper manifolds increases with intensity while the $F=2$ population decreases. The net population flow at the beam center therefore reverses from bright-to-dark to dark-to-bright. This reversal occurs when power broadening activates transitions out of the nominally dark manifold. Once the power-broadened linewidth becomes comparable to the approximately 3-GHz ground-state hyperfine splitting, atoms in $F=2$ can be excited to $F'=2$ or $F'=3$ and subsequently decay into $F=3$. This pathway returns atoms to the principal-transition manifold. We refer to this reversal of the usual population flow as anomalous optical pumping.

As the intensity increases, the power-broadened linewidth first reaches the excited-state splitting and then the ground-state splitting, producing conventional and anomalous optical pumping, respectively. This ordering favors conventional optical pumping at zero magnetic field. We next consider the reversed ordering. An axial magnetic field of 2500 G places the atom in the hyperfine Paschen--Back regime [Fig.~\ref{fig:fig2}(b)], and the laser addresses the $m_J=1/2\rightarrow m_J'=3/2$ transition. In contrast to the zero-field case, the $m_J=-1/2\rightarrow m_J'=1/2$ transition responsible for anomalous optical pumping lies closer to the principal transition than does the $m_J=1/2\rightarrow m_J'=-1/2$ transition responsible for conventional optical pumping. Power broadening therefore activates the anomalous channel first. As shown in Fig.~\ref{fig:fig2}(e), anomalous optical pumping then occurs across most of the beam rather than only at its center, and the calculated population of the nominally dark $m_J=-1/2$ manifold can fall below that of the excited $m_J'=3/2$ manifold. More importantly, a nonzero population difference between the lower and upper states of the principal transition is maintained in the high-intensity region, so the medium retains appreciable absorptive and dispersive responses under strong driving. We test this prediction below using atomic-filter transmission spectra.

\begin{figure}
\includegraphics[width=1\linewidth]{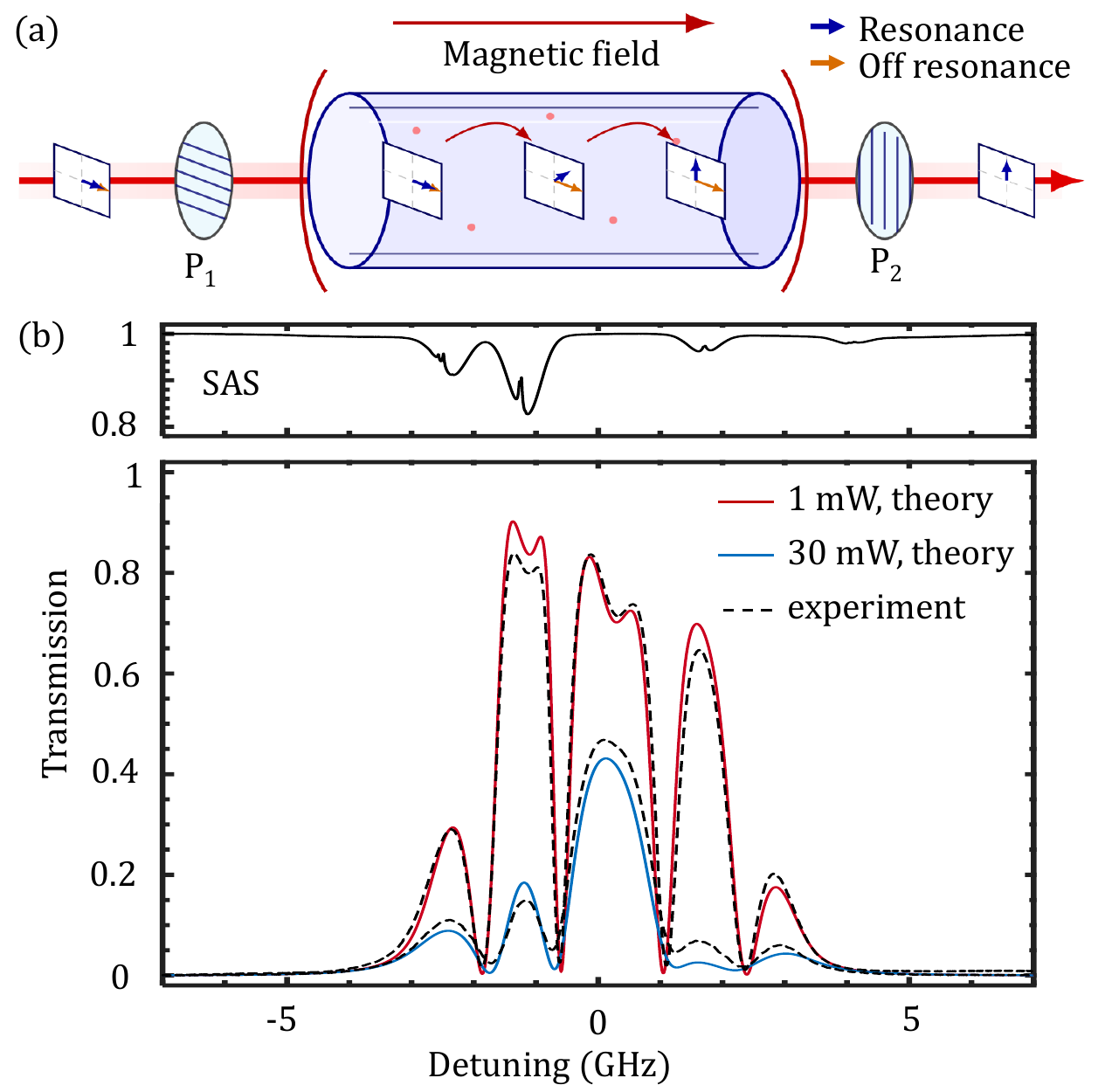}
\caption{\label{fig:fig3}
(a) Operating principle of the Faraday atomic filter. A linearly polarized beam traverses a vapor cell in an axial magnetic field and is transmitted by the crossed analyzer when the polarization rotation approaches $\pi/2$.
(b) Measured saturated-absorption spectrum (SAS; upper panel) and transmission spectra of a $^{85}$Rb filter at 300 G and 50 $^\circ$C (lower panel). The red and blue solid curves are the theoretical spectra at input powers of 1 and 30 mW, respectively; the black dashed curves show the corresponding experimental spectra and nearly overlap at the two powers.
}
\end{figure}

The experiment uses a Faraday anomalous-dispersion optical filter (FADOF), whose operating principle is shown in Fig.~\ref{fig:fig3}(a) \cite{yeh1982dispersive,yin1991theoretical}. Light transmitted by polarizer P1 enters an atomic vapor cell in an axial magnetic field. The two circular polarization components experience different complex refractive indices; the resulting circular birefringence rotates the polarization, while circular dichroism and ordinary absorption attenuate the field. In the absence of appreciable absorption and ellipticity, a rotation of $\pi/2$ gives maximum transmission through the crossed analyzer P2. Because the complex susceptibility depends on the population differences and optical coherences of the relevant transitions, the filter spectrum provides a sensitive probe of optical pumping.

We measure the $^{85}$Rb FADOF transmission spectrum in both the low-field and hyperfine Paschen--Back regimes. In both measurements, the laser beam has a Gaussian $1/e^2$ intensity radius $w=0.15$ mm, and the cylindrical vapor cell has a radius of 10 mm and a length of 30 mm. Throughout, the reported experimental and theoretical transmissions are relative atomic-filter transmissions normalized to exclude losses from the polarizing beam splitters and the vapor cell. In the low-field measurement, a nearly uniform 300-G field is applied, so conventional optical pumping is activated first. At an input power of 1 mW, the cell temperature is set to approximately 50 $^\circ$C to maximize the overall transmission while keeping the Faraday rotation below $\pi$, thereby avoiding additional spectral lobes associated with higher-order rotations. The input power is then increased from 1 to 30 mW. As shown in Fig.~\ref{fig:fig3}(b), the measured and calculated spectra both show the near-resonant peak decreasing from 85$\%$ to 15$\%$ and the central peak decreasing from 80$\%$ to 40$\%$. This behavior is consistent with rapid saturation of the principal transition caused by conventional optical pumping in the low-field regime.

\begin{figure}
\includegraphics[width=1\linewidth]{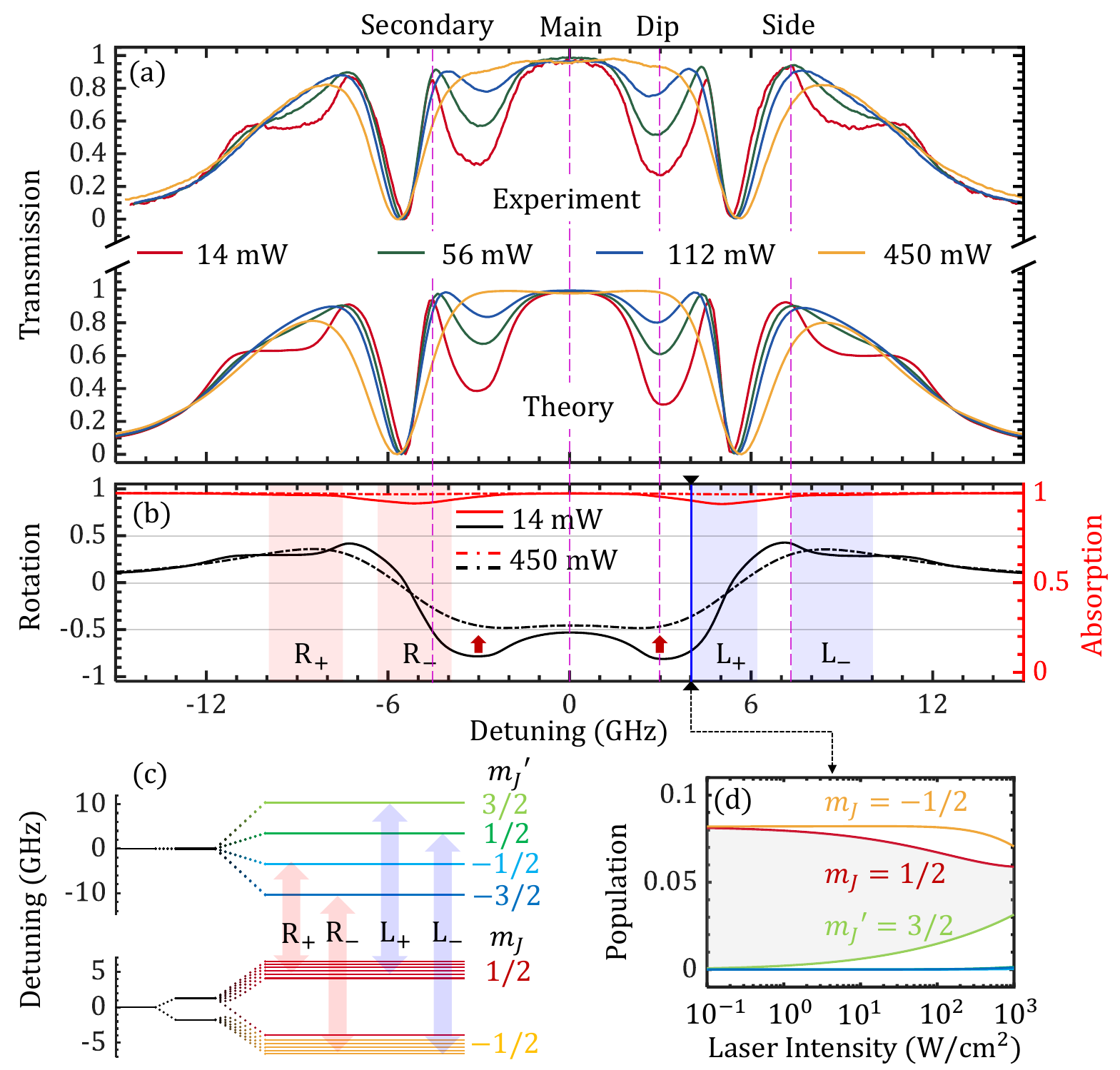}
\caption{\label{fig:fig4} $^{85}$Rb FADOF spectra in the hyperfine Paschen--Back regime. (a) Measured and calculated transmission spectra at 90 $^\circ$C in an axial magnetic field with a mean value of 3700 G, for input powers from 14 to 450 mW. (b) Calculated polarization rotation and absorption spectra. Transmission maxima occur near a $\pi/2$ rotation, whereas the spectral dip is associated with rotation beyond $\pi/2$. (c) The four transition groups $L_+$, $L_-$, $R_+$, and $R_-$ in the hyperfine Paschen--Back regime; their spectral positions are marked in (b). (d) Calculated populations of the relevant Zeeman-state groups as functions of laser intensity at the detuning indicated in (b).}
\end{figure}

In the hyperfine Paschen--Back measurement, the axial magnetic field has a mean value of 3700 G. The finite-length magnet produces an inhomogeneous field ranging from 2900 to 4300 G across the cell, with the atoms remaining in the hyperfine Paschen--Back regime throughout. Using the same optimization criterion as in the low-field measurement, we set the cell temperature to 90 $^\circ$C and measure spectra for input powers from 14 to 450 mW. Notably, calculations and measurements remain in close agreement throughout this range, with the maximum power corresponding to a Gaussian peak intensity $I_0=2P/(\pi w^2)=1.27\times10^3$ W/cm$^2$, or $5.1\times10^5$ times the $^{85}$Rb D2-line saturation intensity of 2.504 mW/cm$^2$ \cite{steck2025rb85}. Figure~\ref{fig:fig4}(a) also shows that the peak transmission remains high across the measured power range. No parameters are fitted: the measured optical power, magnetic field, and cell temperature are used directly in the calculation, with the same model parameters applied at all powers. At 14 mW, the spectrum contains a principal peak, a secondary peak, and two sidebands, with a pronounced dip between the two peaks. The dip progressively fills in with increasing power. At 450 mW, it disappears and the principal and secondary peaks merge into a flat-top profile spanning approximately 7 GHz whose measured and calculated relative transmissions both remain close to 97$\%$. Figures~\ref{fig:fig4}(b)--(d) explain this evolution. In Fig.~\ref{fig:fig4}(b), the transmission maxima coincide with polarization rotation near $\pi/2$, whereas the dip occurs where the rotation exceeds $\pi/2$. Figure~\ref{fig:fig4}(c) identifies the four transition groups $L_+$, $L_-$, $R_+$, and $R_-$. The $L_+$ and $R_-$ groups contain the cycling transitions and have larger dipole matrix elements than the $L_-$ and $R_+$ groups, so they dominate the spectrum. When $L_+$ ($R_-$) is the principal group, $L_-$ ($R_+$) provides the anomalous optical-pumping pathway. Increasing intensity reduces rotations beyond $\pi/2$ through saturation of the principal transition, which fills in the dip. At the same time, anomalous optical pumping replenishes the bright-state population and counteracts saturation, preserving a rotation near $\pi/2$ and a relative transmission of approximately 97$\%$ at 450 mW.

In summary, we have connected the trajectory-dependent internal-state evolution of individual atoms to the spectral response of a thermal ensemble under strong optical driving. The model reveals anomalous optical pumping and shows that hyperfine Paschen--Back splitting can enhance this mechanism, allowing both the measured and calculated relative filter transmission to remain near 97$\%$ at the highest intensity studied. The resulting framework provides a general tool for designing saturation-resistant narrowband atomic filters for spectral selection under intense illumination, including strong-background rejection in free-space optical receivers and filters based on excited-state transitions at telecommunications wavelengths. The framework also enables spatially resolved studies of thermal atoms interacting with intense structured optical fields.

\clearpage

\renewcommand{\theequation}{S\arabic{equation}}
\renewcommand{\theHequation}{S\arabic{equation}}
\setcounter{section}{0}
\setcounter{equation}{0}
\setcounter{figure}{0}
\setcounter{table}{0}

\ifdefined\arxivpreprint\else\onecolumngrid\fi

\begin{center}
{\large\bfseries Supplemental Material for:\\[3pt]
State-Dependent Diffusion and Spectra of Strongly Driven Thermal Atoms}
\end{center}

\ifdefined\arxivpreprint\else\twocolumngrid\fi

\vspace{0.5\baselineskip}
This Supplemental Material provides the technical details underlying the model and spectral calculations in the main text. Section~I derives the diffusion-form master equation from the phase-space kinetic equation, states the moment-closure and adiabatic-elimination assumptions, and gives the accompanying stationary density and heat equations. Section~II specifies the internal-state generator, including its Hamiltonian and dissipative channels, and defines the internal-state parameters used in the calculation. Section~III describes how the resulting spatially resolved atomic state is converted into the susceptibility tensor and the transmission spectrum of the atomic filter.

\section{Derivation of the diffusion-form master equation}

\subsection{Phase-space equation and exact moment hierarchy}

Let $\varrho(t,\mathbf{x},\mathbf{v})$ be the density-matrix-valued distribution of atoms at transverse position $\mathbf{x}$ and velocity $\mathbf{v}$. Its trace is the classical phase-space number density, whereas its matrix elements resolve the internal populations and coherences. Treating the center-of-mass motion semiclassically gives
\begin{equation}
 \partial_t\varrho+\mathbf{v}\cdot\nabla_{\mathbf{x}}\varrho
 +\frac{\mathbf{F}}{m}\cdot\nabla_{\mathbf{v}}\varrho
 =\mathcal{G}(\varrho),
 \label{eq:S1}
\end{equation}
where $\mathbf{F}$ is the transverse optical force and $\mathcal{G}$ is the local internal-state generator, including coherent driving, spontaneous emission, and collisional relaxation \cite{happer1972optical,berman1982collision,firstenberg2013coherent}. Recoil associated with individual spontaneous-emission events is neglected in Eq.~(\ref{eq:S1}).

We define the first three velocity moments by
\begin{equation}
 \bar{\varrho}=\int \varrho\,d^2v,\qquad
 \mathbf{J}=\int \mathbf{v}\varrho\,d^2v,\qquad
 \mathbf{P}=\int \mathbf{v}\mathbf{v}\varrho\,d^2v.
 \label{eq:S2}
\end{equation}
Here $\bar{\varrho}$ is the local density-matrix field, $\mathbf{J}$ is its flux, and $\mathbf{P}$ is the second-moment tensor. In particular, $n=\operatorname{Tr}\bar{\varrho}$ is the particle density and $\boldsymbol{\Phi}_n=\operatorname{Tr}\mathbf{J}$ is the particle current. Integrating Eq.~(\ref{eq:S1}) over velocity and then multiplying it by $\mathbf{v}$ before integration yields the exact relations
\begin{align}
 \partial_t\bar{\varrho}+\nabla\cdot\mathbf{J}
 &=\mathcal{G}(\bar{\varrho}), \label{eq:S3}\\
 \partial_t\mathbf{J}+\nabla\cdot\mathbf{P}
 -\frac{\mathbf{F}}{m}\bar{\varrho}
 &=\mathcal{G}(\mathbf{J}). \label{eq:S4}
\end{align}

\subsection{Maxwellian first-moment closure}

For motion in the two-dimensional transverse plane, the normalized local Maxwell distribution is
\begin{equation}
 f_T(\mathbf{v})=\frac{m}{2\pi k_BT}
 \exp\!\left(-\frac{mv^2}{2k_BT}\right),
 \label{eq:S5}
\end{equation}
which satisfies
\begin{equation}
 \int f_T\,d^2v=1,\qquad
 \int v_i v_j f_T\,d^2v=\frac{k_BT}{m}\delta_{ij}.
 \label{eq:S6}
\end{equation}
When the optical-field variation length $R$ is large compared with the internal relaxation length $v/\Gamma$, higher velocity moments relax rapidly. Retaining the isotropic and first angular moments gives the $P_1$ reconstruction \cite{firstenberg2008theory,finkelstein2023practical}
\begin{equation}
 \varrho(t,\mathbf{x},\mathbf{v})=f_T(\mathbf{v})
 \left[\bar{\varrho}(t,\mathbf{x})
 +\frac{m\mathbf{v}}{k_BT}\cdot\mathbf{J}(t,\mathbf{x})\right].
 \label{eq:S7}
\end{equation}
The coefficient of the anisotropic term is fixed by requiring its first moment to equal $\mathbf{J}$. Direct integration gives
\begin{equation}
 \int\varrho\,d^2v=\bar{\varrho},\qquad
 \int\mathbf{v}\varrho\,d^2v=\mathbf{J},\qquad
 \mathbf{P}=\frac{k_BT}{m}\bar{\varrho}\,\mathbf{I}.
 \label{eq:S8}
\end{equation}
Thus the closure amounts to an isotropic Maxwellian second moment. Substitution into Eqs.~(\ref{eq:S3}) and (\ref{eq:S4}) gives
\begin{align}
 \partial_t\bar{\varrho}+\nabla\cdot\mathbf{J}
 &=\mathcal{G}(\bar{\varrho}), \label{eq:S9}\\
 \partial_t\mathbf{J}
 +\nabla\!\left(\frac{k_BT}{m}\bar{\varrho}\right)
 -\frac{\mathbf{F}}{m}\bar{\varrho}
 &=\mathcal{G}(\mathbf{J}). \label{eq:S10}
\end{align}

\subsection{Adiabatic elimination of the flux}

We next assume that the flux relaxes on a time scale short compared with the evolution of $\bar{\varrho}$. Setting $\partial_t\mathbf{J}=0$ in Eq.~(\ref{eq:S10}) gives
\begin{equation}
 \mathbf{J}=\mathcal{G}_Q^{-1}
 \left[\nabla\!\left(\frac{k_BT}{m}\bar{\varrho}\right)
 -\frac{\mathbf{F}}{m}\bar{\varrho}\right].
 \label{eq:S11}
\end{equation}
Because particle number is conserved, $\mathcal{G}$ contains a nondecaying scalar mode. The symbol $\mathcal{G}_Q^{-1}$ therefore denotes the inverse restricted to the complementary subspace $Q$ of relaxing population and coherence modes; the conserved scalar component is fixed separately by the particle-number equation below. Inserting Eq.~(\ref{eq:S11}) into Eq.~(\ref{eq:S9}) yields
\begin{equation}
 \partial_t\bar{\varrho}
 +\nabla\cdot\left\{\mathcal{G}_Q^{-1}
 \left[\nabla\!\left(\frac{k_BT}{m}\bar{\varrho}\right)
 -\frac{\mathbf{F}}{m}\bar{\varrho}\right]\right\}
 =\mathcal{G}(\bar{\varrho}).
 \label{eq:S12}
\end{equation}
Equation~(\ref{eq:S12}) is the diffusion-form master equation used in the main text. Its transport operator is matrix valued: each relaxing eigenmode of $\mathcal{G}$ is weighted by the inverse of its own relaxation eigenvalue. Consequently, strongly driven population and coherence modes need not share a single diffusion coefficient. This result generalizes spatial density-matrix models based on a prescribed scalar diffusion coefficient \cite{torrey1956bloch,firstenberg2008theory,olsen2011optical}.

In an isothermal, force-free system for which all relevant modes relax at a common momentum-relaxation rate $\Gamma_{\mathrm{tr}}$, one has $\mathcal{G}_Q^{-1}\rightarrow-\Gamma_{\mathrm{tr}}^{-1}$. Equation~(\ref{eq:S12}) then reduces to
\begin{equation}
 \partial_t\bar{\varrho}=\mathcal{G}(\bar{\varrho})
 +D\nabla^2\bar{\varrho},
 \qquad
 D=\frac{k_BT_0}{m\Gamma_{\mathrm{tr}}},
 \label{eq:S13}
\end{equation}
recovering the scalar diffusion model and the Einstein relation for $D$ \cite{torrey1956bloch,firstenberg2008theory}.

\subsection{Particle density and temperature}

Taking the trace of Eqs.~(\ref{eq:S9}) and (\ref{eq:S10}) gives
\begin{align}
 \partial_t n+\nabla\cdot\boldsymbol{\Phi}_n&=0, \label{eq:S14}\\
 \partial_t\boldsymbol{\Phi}_n
 +\nabla\!\left(\frac{k_BT}{m}n\right)
 -\frac{\mathbf{F}}{m}n&=0. \label{eq:S15}
\end{align}
The accompanying energy balance in two transverse dimensions is
\begin{equation}
 \partial_t(nk_BT)
 +\nabla\cdot\left(2k_BT\boldsymbol{\Phi}_n-\kappa\nabla T\right)
 -\mathbf{F}\cdot\boldsymbol{\Phi}_n=Q,
 \label{eq:S16}
\end{equation}
where $Q$ is the optical heating rate per unit volume and $\kappa$ is the thermal conductivity. For the stationary solutions considered in the main text, $\boldsymbol{\Phi}_n=0$, and Eqs.~(\ref{eq:S15}) and (\ref{eq:S16}) become
\begin{align}
 \nabla(k_BTn)&=\mathbf{F}n, \label{eq:S17}\\
 \nabla\cdot(\kappa\nabla T)+Q&=0. \label{eq:S18}
\end{align}
If $T=T_0$ and $\mathbf{F}=-\nabla U$, Eq.~(\ref{eq:S17}) gives $n=n_0\exp[-U/(k_BT_0)]$. For a dilute gas with $\kappa=\kappa_0\sqrt{T/T_0}$, the Kirchhoff transformation converts Eq.~(\ref{eq:S18}) into
\begin{equation}
 \nabla^2(T^{3/2})=-\frac{3\sqrt{T_0}}{2\kappa_0}Q,
 \qquad T|_{\partial\Omega}=T_0.
 \label{eq:S19}
\end{equation}
Together, Eqs.~(\ref{eq:S12}), (\ref{eq:S17}), and (\ref{eq:S18}) form the closed stationary model for the internal state, density, and temperature.

\section{Internal-state dynamics}

\subsection{Hamiltonian and dissipative channels}

Let $\ket{g_j}$ and $\ket{e_i}$ denote magnetic-field-dependent ground- and excited-state eigenvectors with angular frequencies $\omega_j^g$ and $\omega_i^e$, respectively. In the rotating-wave approximation and a frame rotating at the laser frequency $\omega$, the local Hamiltonian is \cite{cohentannoudji1992atom,auzinsh2010optically}
\begin{align}
 \frac{H}{\hbar}={}&\sum_i(\omega_i^e-\omega)\ket{e_i}\bra{e_i}
 +\sum_j\omega_j^g\ket{g_j}\bra{g_j}\nonumber\\
 &+\frac{1}{2}\sum_{ij}\left(
 \Omega_{ij}\ket{e_i}\bra{g_j}
 +\Omega_{ij}^{*}\ket{g_j}\bra{e_i}\right).
 \label{eq:S20}
\end{align}
Spontaneous decay from $\ket{e_i}$ to $\ket{g_j}$ is specified by the Lindblad jump operator
\begin{equation}
 L_{ij}^{(\mathrm{sp})}=\sqrt{\Gamma_{ij}}\ket{g_j}\bra{e_i}.
 \label{eq:S21}
\end{equation}
Additional collisional population transfer and dephasing are included through jump operators of the forms
\begin{equation}
 L_{ab}^{(\mathrm{col})}=\sqrt{\gamma_{ab}}\ket{a}\bra{b},
 \qquad
 L_a^{(\phi)}=\sqrt{\gamma_a^{(\phi)}}\ket{a}\bra{a},
 \label{eq:S22}
\end{equation}
with rates appropriate to the vapor-cell conditions. For an arbitrary internal-state operator $X$, define
\begin{equation}
 \mathcal{D}[L]X=LXL^\dagger-\frac{1}{2}
 \left(L^\dagger LX+XL^\dagger L\right).
 \label{eq:S23}
\end{equation}
The local generator appearing in Sec.~I is then
\begin{align}
 \mathcal{G}(X)={}&-\frac{i}{\hbar}[H,X]
 +\sum_{ij}\mathcal{D}[L_{ij}^{(\mathrm{sp})}]X \nonumber\\
 &+\sum_{ab}\mathcal{D}[L_{ab}^{(\mathrm{col})}]X
 +\sum_a\mathcal{D}[L_a^{(\phi)}]X .
 \label{eq:S24}
\end{align}
This superoperator acts on both the zeroth moment $\bar{\varrho}$ and the flux moment $\mathbf{J}$ in Eqs.~(\ref{eq:S9}) and (\ref{eq:S10}). The spectral calculation therefore uses the diffusion-form equation rather than a separate trajectory-time master equation.

\subsection{Dipole matrix elements, Rabi frequencies, and decay rates}

We use the electric-dipole operator $\mathbf{d}$ and its spherical components $d_q$ ($q=0,\pm1$). The complex field amplitude $\boldsymbol{\mathcal{E}}$ is defined by
$\mathbf{E}(\mathbf{r},t)=\tfrac{1}{2}[\boldsymbol{\mathcal{E}}(\mathbf{r})e^{-i\omega t}+\mathrm{c.c.}]$.
With the spherical-tensor scalar-product convention
$\mathbf{d}\cdot\boldsymbol{\mathcal{E}}=\sum_q(-1)^q d_q\mathcal{E}_{-q}$, the transition matrix element and Rabi frequency are
\begin{align}
 d_{ij}^{(q)}&=\bra{e_i}d_q\ket{g_j}, \label{eq:S25}\\
 \Omega_{ij}&=-\frac{1}{\hbar}\sum_{q=-1}^{1}
 (-1)^q\mathcal{E}_{-q}d_{ij}^{(q)}.
 \label{eq:S26}
\end{align}
The overall minus sign is inherited from the interaction $-\mathbf{d}\cdot\mathbf{E}$ and may be absorbed into the optical phase.

The ground- and excited-state eigenvectors are obtained by diagonalizing the hyperfine--Zeeman Hamiltonian using the $^{85}$Rb atomic constants in Ref.~\cite{steck2025rb85}. With $J$ and $J'$ fixed by the ground and excited fine-structure manifolds, respectively, we abbreviate the coupled basis $\ket{(JI)Fm_F}$ as $\ket{F,m_F}$ and write
\begin{align}
 \ket{g_j}&=\sum_{Fm_F}C_{Fm_F}^{(g_j)}\ket{F,m_F},\nonumber\\
 \ket{e_i}&=\sum_{F'm_F'}C_{F'm_F'}^{(e_i)}
 \ket{F',m_F'} .
 \label{eq:S27}
\end{align}
Consequently,
\begin{align}
 d_{ij}^{(q)}={}&
 \sum_{F'm_F'}\sum_{Fm_F}
 \left[C_{F'm_F'}^{(e_i)}\right]^*
 C_{Fm_F}^{(g_j)}\nonumber\\
 &\times\bra{F',m_F'}d_q\ket{F,m_F}.
 \label{eq:S28}
\end{align}
For compactness, we use the notation
\begin{align*}
d_{F'm_F',Fm_F}^{(q)}&=\bra{F',m_F'}d_q\ket{F,m_F},\\
\mathcal{D}_{F'F}&=\langle F'\|\mathbf{d}\|F\rangle,\\
\mathcal{D}_{J'J}&=\langle J'\|\mathbf{d}\|J\rangle.
\end{align*}
For the phase convention used here, the Wigner--Eckart theorem gives
\begin{align}
 d_{F'm_F',Fm_F}^{(q)}={}&(-1)^{F'-m_F'}
 \begin{pmatrix}
 F'&1&F\\[-2pt]
 -m_F'&q&m_F
 \end{pmatrix}\nonumber\\
 &\quad\times
 \mathcal{D}_{F'F},
 \label{eq:S29}\\
 \mathcal{D}_{F'F}={}&(-1)^{J'+I+F+1}
 \sqrt{(2F'+1)(2F+1)}\nonumber\\
 &\quad\times
 \left\{\begin{matrix}
 J'&F'&I\\
 F&J&1
 \end{matrix}\right\}
 \mathcal{D}_{J'J}.
 \label{eq:S30}
\end{align}
The $3j$ symbol enforces $m_F'=m_F+q$; hence $q=0$ corresponds to $\pi$ excitation and $q=\pm1$ to $\sigma_\pm$ excitation.

The partial spontaneous-emission rate from $\ket{e_i}$ to $\ket{g_j}$ is
\begin{equation}
 \Gamma_{ij}=
 \frac{\omega_{ij}^3}{3\pi\varepsilon_0\hbar c^3}
 \sum_{q=-1}^{1}\left|\bra{g_j}d_q\ket{e_i}\right|^2,
 \label{eq:S31}
\end{equation}
where $\omega_{ij}=\omega_i^e-\omega_j^g>0$. The sum over $q$ accounts for all emitted spherical polarizations. If the set of lower states is complete, the total radiative width is $\Gamma_i=\sum_j\Gamma_{ij}$. Neglecting the small hyperfine dependence of the optical frequency, the fine-structure lifetime is related to the electronic reduced matrix element by
\begin{equation}
 \frac{1}{\tau}=
 \frac{\omega_0^3}{3\pi\varepsilon_0\hbar c^3}
 \frac{|\mathcal{D}_{J'J}|^2}{2J'+1}.
 \label{eq:S32}
\end{equation}

\section{Atomic-filter transmission spectrum}

Once the spatially resolved density-matrix field has been obtained from Eq.~(\ref{eq:S12}), its optical coherences determine the polarization and hence the effective susceptibility tensor used to propagate the optical field \cite{yeh1982dispersive,yin1991theoretical}. The macroscopic polarization is
\begin{equation}
 \mathbf{P}=\varepsilon_0\boldsymbol{\chi}\mathbf{E},
 \label{eq:S33}
\end{equation}
and the monochromatic field obeys
\begin{equation}
 \nabla^2\mathbf{E}-\frac{1}{c^2}\frac{\partial^2\mathbf{E}}{\partial t^2}
 =\mu_0\frac{\partial^2\mathbf{P}}{\partial t^2}.
 \label{eq:S34}
\end{equation}
For propagation along the cell axis $z$, write $\mathbf{E}(\mathbf{r},t)=\mathbf{E}(z)e^{-i\omega t}$ and define the relative permittivity tensor $\boldsymbol{\varepsilon}=\mathbf{I}+\boldsymbol{\chi}$. The condition $\nabla\cdot\mathbf{D}=0$ gives
\begin{equation}
 E_z=-\frac{\varepsilon_{zx}E_x+\varepsilon_{zy}E_y}{\varepsilon_{zz}}.
 \label{eq:S35}
\end{equation}
Eliminating $E_z$ leaves the transverse wave equation
\begin{equation}
 \frac{d^2\mathbf{E}_{\perp}}{dz^2}+\mathbf{K}^2\mathbf{E}_{\perp}=0,
 \qquad
 \mathbf{E}_{\perp}=\begin{pmatrix}E_x\\E_y\end{pmatrix},
 \label{eq:S36}
\end{equation}
where
\begin{equation}
 \mathbf{K}^2=\frac{\omega^2}{c^2}
 \left[
 \begin{pmatrix}
 \varepsilon_{xx}&\varepsilon_{xy}\\
 \varepsilon_{yx}&\varepsilon_{yy}
 \end{pmatrix}
 -\frac{1}{\varepsilon_{zz}}
 \begin{pmatrix}\varepsilon_{xz}\\\varepsilon_{yz}\end{pmatrix}
 \begin{pmatrix}\varepsilon_{zx}&\varepsilon_{zy}\end{pmatrix}
 \right].
 \label{eq:S37}
\end{equation}
If $\mathbf{K}^2=\mathbf{V}\mathbf{k}^2\mathbf{V}^{-1}$, with the forward-propagating square-root branch chosen for each eigenvalue, propagation through a cell of length $L$ gives
\begin{equation}
 \mathbf{E}_{\perp}(L)=\mathbf{V}e^{i\mathbf{k}L}\mathbf{V}^{-1}
 \mathbf{E}_{\perp}(0).
 \label{eq:S38}
\end{equation}
Let $\mathbf{P}_1$ and $\mathbf{P}_2$ be the Jones matrices of the entrance polarizer and crossed analyzer. The output field and intensity transmission are
\begin{align}
 \mathbf{E}_{\mathrm{out}}&=\mathbf{P}_2\mathbf{V}e^{i\mathbf{k}L}
 \mathbf{V}^{-1}\mathbf{P}_1\mathbf{E}_{\mathrm{in}}, \label{eq:S39}\\
 \mathcal{T}(\omega)&=\frac{\mathbf{E}_{\mathrm{out}}^{\dagger}
 \mathbf{E}_{\mathrm{out}}}{\mathbf{E}_{\mathrm{in}}^{\dagger}
 \mathbf{E}_{\mathrm{in}}}.
 \label{eq:S40}
\end{align}
The axial Doppler average is performed separately for each velocity class when constructing $\boldsymbol{\chi}$, whereas transverse transport is already contained in the solution of Eq.~(\ref{eq:S12}).

\bibliography{manuscript}

\begin{thebibliography}{43}%
\makeatletter
\providecommand \@ifxundefined [1]{%
 \@ifx{#1\undefined}
}%
\providecommand \@ifnum [1]{%
 \ifnum #1\expandafter \@firstoftwo
 \else \expandafter \@secondoftwo
 \fi
}%
\providecommand \@ifx [1]{%
 \ifx #1\expandafter \@firstoftwo
 \else \expandafter \@secondoftwo
 \fi
}%
\providecommand \natexlab [1]{#1}%
\providecommand \enquote  [1]{``#1''}%
\providecommand \bibnamefont  [1]{#1}%
\providecommand \bibfnamefont [1]{#1}%
\providecommand \citenamefont [1]{#1}%
\providecommand \href@noop [0]{\@secondoftwo}%
\providecommand \href [0]{\begingroup \@sanitize@url \@href}%
\providecommand \@href[1]{\@@startlink{#1}\@@href}%
\providecommand \@@href[1]{\endgroup#1\@@endlink}%
\providecommand \@sanitize@url [0]{\catcode `\\12\catcode `\$12\catcode
  `\&12\catcode `\#12\catcode `\^12\catcode `\_12\catcode `\%12\relax}%
\providecommand \@@startlink[1]{}%
\providecommand \@@endlink[0]{}%
\providecommand \url  [0]{\begingroup\@sanitize@url \@url }%
\providecommand \@url [1]{\endgroup\@href {#1}{\urlprefix }}%
\providecommand \urlprefix  [0]{URL }%
\providecommand \Eprint [0]{\href }%
\providecommand \doibase [0]{https://doi.org/}%
\providecommand \selectlanguage [0]{\@gobble}%
\providecommand \bibinfo  [0]{\@secondoftwo}%
\providecommand \bibfield  [0]{\@secondoftwo}%
\providecommand \translation [1]{[#1]}%
\providecommand \BibitemOpen [0]{}%
\providecommand \bibitemStop [0]{}%
\providecommand \bibitemNoStop [0]{.\EOS\space}%
\providecommand \EOS [0]{\spacefactor3000\relax}%
\providecommand \BibitemShut  [1]{\csname bibitem#1\endcsname}%
\let\auto@bib@innerbib\@empty
\bibitem [{\citenamefont {Hilbert}(1902)}]{hilbert1902mathematical}%
  \BibitemOpen
  \bibfield  {author} {\bibinfo {author} {\bibfnamefont {D.}~\bibnamefont
  {Hilbert}},\ }\href {https://doi.org/10.1090/S0002-9904-1902-00923-3}
  {\bibfield  {journal} {\bibinfo  {journal} {Bull. Amer. Math. Soc.}\ }\textbf
  {\bibinfo {volume} {8}},\ \bibinfo {pages} {437} (\bibinfo {year}
  {1902})}\BibitemShut {NoStop}%
\bibitem [{\citenamefont {Deng}\ \emph {et~al.}(2025)\citenamefont {Deng},
  \citenamefont {Hani},\ and\ \citenamefont {Ma}}]{deng2025hilbert}%
  \BibitemOpen
  \bibfield  {author} {\bibinfo {author} {\bibfnamefont {Y.}~\bibnamefont
  {Deng}}, \bibinfo {author} {\bibfnamefont {Z.}~\bibnamefont {Hani}},\ and\
  \bibinfo {author} {\bibfnamefont {X.}~\bibnamefont {Ma}},\ }\href
  {https://doi.org/10.48550/arXiv.2503.01800} {\bibinfo {title} {Hilbert's
  sixth problem: derivation of fluid equations via boltzmann's kinetic theory}}
  (\bibinfo {year} {2025}),\ \Eprint {https://arxiv.org/abs/2503.01800}
  {arXiv:2503.01800} \BibitemShut {NoStop}%
\bibitem [{\citenamefont {Cohen-Tannoudji}\ \emph {et~al.}(1992)\citenamefont
  {Cohen-Tannoudji}, \citenamefont {Dupont-Roc},\ and\ \citenamefont
  {Grynberg}}]{cohentannoudji1992atom}%
  \BibitemOpen
  \bibfield  {author} {\bibinfo {author} {\bibfnamefont {C.}~\bibnamefont
  {Cohen-Tannoudji}}, \bibinfo {author} {\bibfnamefont {J.}~\bibnamefont
  {Dupont-Roc}},\ and\ \bibinfo {author} {\bibfnamefont {G.}~\bibnamefont
  {Grynberg}},\ }\href {https://doi.org/10.1002/9783527617197} {\emph {\bibinfo
  {title} {Atom-Photon Interactions: Basic Processes and Applications}}}\
  (\bibinfo  {publisher} {Wiley},\ \bibinfo {address} {New York},\ \bibinfo
  {year} {1992})\BibitemShut {NoStop}%
\bibitem [{\citenamefont {Auzinsh}\ \emph {et~al.}(2010)\citenamefont
  {Auzinsh}, \citenamefont {Budker},\ and\ \citenamefont
  {Rochester}}]{auzinsh2010optically}%
  \BibitemOpen
  \bibfield  {author} {\bibinfo {author} {\bibfnamefont {M.}~\bibnamefont
  {Auzinsh}}, \bibinfo {author} {\bibfnamefont {D.}~\bibnamefont {Budker}},\
  and\ \bibinfo {author} {\bibfnamefont {S.~M.}\ \bibnamefont {Rochester}},\
  }\href {https://books.google.com/books?vid=ISBN9780199565122} {\emph
  {\bibinfo {title} {Optically Polarized Atoms: Understanding Light-Atom
  Interactions}}}\ (\bibinfo  {publisher} {Oxford University Press},\ \bibinfo
  {year} {2010})\BibitemShut {NoStop}%
\bibitem [{\citenamefont {Ludlow}\ \emph {et~al.}(2015)\citenamefont {Ludlow},
  \citenamefont {Boyd}, \citenamefont {Ye}, \citenamefont {Peik},\ and\
  \citenamefont {Schmidt}}]{ludlow2015optical}%
  \BibitemOpen
  \bibfield  {author} {\bibinfo {author} {\bibfnamefont {A.~D.}\ \bibnamefont
  {Ludlow}}, \bibinfo {author} {\bibfnamefont {M.~M.}\ \bibnamefont {Boyd}},
  \bibinfo {author} {\bibfnamefont {J.}~\bibnamefont {Ye}}, \bibinfo {author}
  {\bibfnamefont {E.}~\bibnamefont {Peik}},\ and\ \bibinfo {author}
  {\bibfnamefont {P.~O.}\ \bibnamefont {Schmidt}},\ }\href
  {https://doi.org/10.1103/RevModPhys.87.637} {\bibfield  {journal} {\bibinfo
  {journal} {Rev. Mod. Phys.}\ }\textbf {\bibinfo {volume} {87}},\ \bibinfo
  {pages} {637} (\bibinfo {year} {2015})}\BibitemShut {NoStop}%
\bibitem [{\citenamefont {Takamoto}\ \emph {et~al.}(2005)\citenamefont
  {Takamoto}, \citenamefont {Hong}, \citenamefont {Higashi},\ and\
  \citenamefont {Katori}}]{takamoto2005optical}%
  \BibitemOpen
  \bibfield  {author} {\bibinfo {author} {\bibfnamefont {M.}~\bibnamefont
  {Takamoto}}, \bibinfo {author} {\bibfnamefont {F.-L.}\ \bibnamefont {Hong}},
  \bibinfo {author} {\bibfnamefont {R.}~\bibnamefont {Higashi}},\ and\ \bibinfo
  {author} {\bibfnamefont {H.}~\bibnamefont {Katori}},\ }\href
  {https://doi.org/10.1038/nature03541} {\bibfield  {journal} {\bibinfo
  {journal} {Nature}\ }\textbf {\bibinfo {volume} {435}},\ \bibinfo {pages}
  {321} (\bibinfo {year} {2005})}\BibitemShut {NoStop}%
\bibitem [{\citenamefont {Knappe}\ \emph {et~al.}(2004)\citenamefont {Knappe},
  \citenamefont {Shah}, \citenamefont {Schwindt}, \citenamefont {Hollberg},
  \citenamefont {Kitching}, \citenamefont {Liew},\ and\ \citenamefont
  {Moreland}}]{knappe2004microfabricated}%
  \BibitemOpen
  \bibfield  {author} {\bibinfo {author} {\bibfnamefont {S.}~\bibnamefont
  {Knappe}}, \bibinfo {author} {\bibfnamefont {V.}~\bibnamefont {Shah}},
  \bibinfo {author} {\bibfnamefont {P.~D.~D.}\ \bibnamefont {Schwindt}},
  \bibinfo {author} {\bibfnamefont {L.}~\bibnamefont {Hollberg}}, \bibinfo
  {author} {\bibfnamefont {J.}~\bibnamefont {Kitching}}, \bibinfo {author}
  {\bibfnamefont {L.-A.}\ \bibnamefont {Liew}},\ and\ \bibinfo {author}
  {\bibfnamefont {J.}~\bibnamefont {Moreland}},\ }\href
  {https://doi.org/10.1063/1.1787942} {\bibfield  {journal} {\bibinfo
  {journal} {Appl. Phys. Lett.}\ }\textbf {\bibinfo {volume} {85}},\ \bibinfo
  {pages} {1460} (\bibinfo {year} {2004})}\BibitemShut {NoStop}%
\bibitem [{\citenamefont {Kitching}(2018)}]{kitching2018chip}%
  \BibitemOpen
  \bibfield  {author} {\bibinfo {author} {\bibfnamefont {J.}~\bibnamefont
  {Kitching}},\ }\href {https://doi.org/10.1063/1.5026238} {\bibfield
  {journal} {\bibinfo  {journal} {Appl. Phys. Rev.}\ }\textbf {\bibinfo
  {volume} {5}},\ \bibinfo {pages} {031302} (\bibinfo {year}
  {2018})}\BibitemShut {NoStop}%
\bibitem [{\citenamefont {Budker}\ and\ \citenamefont
  {Romalis}(2007)}]{budker2007optical}%
  \BibitemOpen
  \bibfield  {author} {\bibinfo {author} {\bibfnamefont {D.}~\bibnamefont
  {Budker}}\ and\ \bibinfo {author} {\bibfnamefont {M.}~\bibnamefont
  {Romalis}},\ }\href {https://doi.org/10.1038/nphys566} {\bibfield  {journal}
  {\bibinfo  {journal} {Nat. Phys.}\ }\textbf {\bibinfo {volume} {3}},\
  \bibinfo {pages} {227} (\bibinfo {year} {2007})}\BibitemShut {NoStop}%
\bibitem [{\citenamefont {Kominis}\ \emph {et~al.}(2003)\citenamefont
  {Kominis}, \citenamefont {Kornack}, \citenamefont {Allred},\ and\
  \citenamefont {Romalis}}]{kominis2003subfemtotesla}%
  \BibitemOpen
  \bibfield  {author} {\bibinfo {author} {\bibfnamefont {I.~K.}\ \bibnamefont
  {Kominis}}, \bibinfo {author} {\bibfnamefont {T.~W.}\ \bibnamefont
  {Kornack}}, \bibinfo {author} {\bibfnamefont {J.~C.}\ \bibnamefont
  {Allred}},\ and\ \bibinfo {author} {\bibfnamefont {M.~V.}\ \bibnamefont
  {Romalis}},\ }\href {https://doi.org/10.1038/nature01484} {\bibfield
  {journal} {\bibinfo  {journal} {Nature}\ }\textbf {\bibinfo {volume} {422}},\
  \bibinfo {pages} {596} (\bibinfo {year} {2003})}\BibitemShut {NoStop}%
\bibitem [{\citenamefont {Wieman}\ and\ \citenamefont
  {H{\"a}nsch}(1976)}]{wieman1976doppler}%
  \BibitemOpen
  \bibfield  {author} {\bibinfo {author} {\bibfnamefont {C.}~\bibnamefont
  {Wieman}}\ and\ \bibinfo {author} {\bibfnamefont {T.~W.}\ \bibnamefont
  {H{\"a}nsch}},\ }\href {https://doi.org/10.1103/PhysRevLett.36.1170}
  {\bibfield  {journal} {\bibinfo  {journal} {Phys. Rev. Lett.}\ }\textbf
  {\bibinfo {volume} {36}},\ \bibinfo {pages} {1170} (\bibinfo {year}
  {1976})}\BibitemShut {NoStop}%
\bibitem [{\citenamefont {Yang}\ \emph {et~al.}(2011)\citenamefont {Yang},
  \citenamefont {Wang}, \citenamefont {Lu}, \citenamefont {Li}, \citenamefont
  {Hua}, \citenamefont {Xu},\ and\ \citenamefont {Chen}}]{yang2011power}%
  \BibitemOpen
  \bibfield  {author} {\bibinfo {author} {\bibfnamefont {Z.}~\bibnamefont
  {Yang}}, \bibinfo {author} {\bibfnamefont {H.}~\bibnamefont {Wang}}, \bibinfo
  {author} {\bibfnamefont {Q.}~\bibnamefont {Lu}}, \bibinfo {author}
  {\bibfnamefont {Y.}~\bibnamefont {Li}}, \bibinfo {author} {\bibfnamefont
  {W.}~\bibnamefont {Hua}}, \bibinfo {author} {\bibfnamefont {X.}~\bibnamefont
  {Xu}},\ and\ \bibinfo {author} {\bibfnamefont {J.}~\bibnamefont {Chen}},\
  }\href {https://doi.org/10.1364/JOSAB.28.001353} {\bibfield  {journal}
  {\bibinfo  {journal} {J. Opt. Soc. Am. B}\ }\textbf {\bibinfo {volume}
  {28}},\ \bibinfo {pages} {1353} (\bibinfo {year} {2011})}\BibitemShut
  {NoStop}%
\bibitem [{\citenamefont {Krupke}(2012)}]{krupke2012dpal}%
  \BibitemOpen
  \bibfield  {author} {\bibinfo {author} {\bibfnamefont {W.~F.}\ \bibnamefont
  {Krupke}},\ }\href {https://doi.org/10.1016/j.pquantelec.2011.09.001}
  {\bibfield  {journal} {\bibinfo  {journal} {Prog. Quantum Electron.}\
  }\textbf {\bibinfo {volume} {36}},\ \bibinfo {pages} {4} (\bibinfo {year}
  {2012})}\BibitemShut {NoStop}%
\bibitem [{\citenamefont {Hau}\ \emph {et~al.}(1999)\citenamefont {Hau},
  \citenamefont {Harris}, \citenamefont {Dutton},\ and\ \citenamefont
  {Behroozi}}]{hau1999light}%
  \BibitemOpen
  \bibfield  {author} {\bibinfo {author} {\bibfnamefont {L.~V.}\ \bibnamefont
  {Hau}}, \bibinfo {author} {\bibfnamefont {S.~E.}\ \bibnamefont {Harris}},
  \bibinfo {author} {\bibfnamefont {Z.}~\bibnamefont {Dutton}},\ and\ \bibinfo
  {author} {\bibfnamefont {C.~H.}\ \bibnamefont {Behroozi}},\ }\href
  {https://doi.org/10.1038/17561} {\bibfield  {journal} {\bibinfo  {journal}
  {Nature}\ }\textbf {\bibinfo {volume} {397}},\ \bibinfo {pages} {594}
  (\bibinfo {year} {1999})}\BibitemShut {NoStop}%
\bibitem [{\citenamefont {Kash}\ \emph {et~al.}(1999)\citenamefont {Kash},
  \citenamefont {Sautenkov}, \citenamefont {Zibrov}, \citenamefont {Hollberg},
  \citenamefont {Welch}, \citenamefont {Lukin}, \citenamefont {Rostovtsev},
  \citenamefont {Fry},\ and\ \citenamefont {Scully}}]{kash1999ultraslow}%
  \BibitemOpen
  \bibfield  {author} {\bibinfo {author} {\bibfnamefont {M.~M.}\ \bibnamefont
  {Kash}}, \bibinfo {author} {\bibfnamefont {V.~A.}\ \bibnamefont {Sautenkov}},
  \bibinfo {author} {\bibfnamefont {A.~S.}\ \bibnamefont {Zibrov}}, \bibinfo
  {author} {\bibfnamefont {L.}~\bibnamefont {Hollberg}}, \bibinfo {author}
  {\bibfnamefont {G.~R.}\ \bibnamefont {Welch}}, \bibinfo {author}
  {\bibfnamefont {M.~D.}\ \bibnamefont {Lukin}}, \bibinfo {author}
  {\bibfnamefont {Y.}~\bibnamefont {Rostovtsev}}, \bibinfo {author}
  {\bibfnamefont {E.~S.}\ \bibnamefont {Fry}},\ and\ \bibinfo {author}
  {\bibfnamefont {M.~O.}\ \bibnamefont {Scully}},\ }\href
  {https://doi.org/10.1103/PhysRevLett.82.5229} {\bibfield  {journal} {\bibinfo
   {journal} {Phys. Rev. Lett.}\ }\textbf {\bibinfo {volume} {82}},\ \bibinfo
  {pages} {5229} (\bibinfo {year} {1999})}\BibitemShut {NoStop}%
\bibitem [{\citenamefont {Fleischhauer}\ and\ \citenamefont
  {Lukin}(2000)}]{fleischhauer2000dark}%
  \BibitemOpen
  \bibfield  {author} {\bibinfo {author} {\bibfnamefont {M.}~\bibnamefont
  {Fleischhauer}}\ and\ \bibinfo {author} {\bibfnamefont {M.~D.}\ \bibnamefont
  {Lukin}},\ }\href {https://doi.org/10.1103/PhysRevLett.84.5094} {\bibfield
  {journal} {\bibinfo  {journal} {Phys. Rev. Lett.}\ }\textbf {\bibinfo
  {volume} {84}},\ \bibinfo {pages} {5094} (\bibinfo {year}
  {2000})}\BibitemShut {NoStop}%
\bibitem [{\citenamefont {Julsgaard}\ \emph {et~al.}(2004)\citenamefont
  {Julsgaard}, \citenamefont {Sherson}, \citenamefont {Cirac}, \citenamefont
  {Fiur{\'a}{\v{s}}ek},\ and\ \citenamefont
  {Polzik}}]{julsgaard2004experimental}%
  \BibitemOpen
  \bibfield  {author} {\bibinfo {author} {\bibfnamefont {B.}~\bibnamefont
  {Julsgaard}}, \bibinfo {author} {\bibfnamefont {J.}~\bibnamefont {Sherson}},
  \bibinfo {author} {\bibfnamefont {J.~I.}\ \bibnamefont {Cirac}}, \bibinfo
  {author} {\bibfnamefont {J.}~\bibnamefont {Fiur{\'a}{\v{s}}ek}},\ and\
  \bibinfo {author} {\bibfnamefont {E.~S.}\ \bibnamefont {Polzik}},\ }\href
  {https://doi.org/10.1038/nature03064} {\bibfield  {journal} {\bibinfo
  {journal} {Nature}\ }\textbf {\bibinfo {volume} {432}},\ \bibinfo {pages}
  {482} (\bibinfo {year} {2004})}\BibitemShut {NoStop}%
\bibitem [{\citenamefont {Menders}\ \emph {et~al.}(1991)\citenamefont
  {Menders}, \citenamefont {Benson}, \citenamefont {Bloom}, \citenamefont
  {Liu},\ and\ \citenamefont {Korevaar}}]{menders1991ultranarrow}%
  \BibitemOpen
  \bibfield  {author} {\bibinfo {author} {\bibfnamefont {J.}~\bibnamefont
  {Menders}}, \bibinfo {author} {\bibfnamefont {K.}~\bibnamefont {Benson}},
  \bibinfo {author} {\bibfnamefont {S.~H.}\ \bibnamefont {Bloom}}, \bibinfo
  {author} {\bibfnamefont {C.~S.}\ \bibnamefont {Liu}},\ and\ \bibinfo {author}
  {\bibfnamefont {E.}~\bibnamefont {Korevaar}},\ }\href
  {https://doi.org/10.1364/OL.16.000846} {\bibfield  {journal} {\bibinfo
  {journal} {Opt. Lett.}\ }\textbf {\bibinfo {volume} {16}},\ \bibinfo {pages}
  {846} (\bibinfo {year} {1991})}\BibitemShut {NoStop}%
\bibitem [{\citenamefont {Tang}\ \emph {et~al.}(1995)\citenamefont {Tang},
  \citenamefont {Wang}, \citenamefont {Li}, \citenamefont {Zhang},
  \citenamefont {Gan}, \citenamefont {Duan}, \citenamefont {Kong},\ and\
  \citenamefont {Zheng}}]{tang1995experimental}%
  \BibitemOpen
  \bibfield  {author} {\bibinfo {author} {\bibfnamefont {J.}~\bibnamefont
  {Tang}}, \bibinfo {author} {\bibfnamefont {Q.}~\bibnamefont {Wang}}, \bibinfo
  {author} {\bibfnamefont {Y.}~\bibnamefont {Li}}, \bibinfo {author}
  {\bibfnamefont {L.}~\bibnamefont {Zhang}}, \bibinfo {author} {\bibfnamefont
  {J.}~\bibnamefont {Gan}}, \bibinfo {author} {\bibfnamefont {M.}~\bibnamefont
  {Duan}}, \bibinfo {author} {\bibfnamefont {J.}~\bibnamefont {Kong}},\ and\
  \bibinfo {author} {\bibfnamefont {L.}~\bibnamefont {Zheng}},\ }\href
  {https://doi.org/10.1364/AO.34.002619} {\bibfield  {journal} {\bibinfo
  {journal} {Appl. Opt.}\ }\textbf {\bibinfo {volume} {34}},\ \bibinfo {pages}
  {2619} (\bibinfo {year} {1995})}\BibitemShut {NoStop}%
\bibitem [{\citenamefont {Fricke-Begemann}\ \emph {et~al.}(2002)\citenamefont
  {Fricke-Begemann}, \citenamefont {Alpers},\ and\ \citenamefont
  {H{\"o}ffner}}]{frickebegemann2002daylight}%
  \BibitemOpen
  \bibfield  {author} {\bibinfo {author} {\bibfnamefont {C.}~\bibnamefont
  {Fricke-Begemann}}, \bibinfo {author} {\bibfnamefont {M.}~\bibnamefont
  {Alpers}},\ and\ \bibinfo {author} {\bibfnamefont {J.}~\bibnamefont
  {H{\"o}ffner}},\ }\href {https://doi.org/10.1364/OL.27.001932} {\bibfield
  {journal} {\bibinfo  {journal} {Opt. Lett.}\ }\textbf {\bibinfo {volume}
  {27}},\ \bibinfo {pages} {1932} (\bibinfo {year} {2002})}\BibitemShut
  {NoStop}%
\bibitem [{\citenamefont {Yin}\ \emph {et~al.}(2024)\citenamefont {Yin},
  \citenamefont {Zhan}, \citenamefont {Tang}, \citenamefont {Ge}, \citenamefont
  {Chen}, \citenamefont {Li}, \citenamefont {Wu},\ and\ \citenamefont
  {Luo}}]{yin2024ghost}%
  \BibitemOpen
  \bibfield  {author} {\bibinfo {author} {\bibfnamefont {L.}~\bibnamefont
  {Yin}}, \bibinfo {author} {\bibfnamefont {H.}~\bibnamefont {Zhan}}, \bibinfo
  {author} {\bibfnamefont {W.}~\bibnamefont {Tang}}, \bibinfo {author}
  {\bibfnamefont {H.}~\bibnamefont {Ge}}, \bibinfo {author} {\bibfnamefont
  {L.}~\bibnamefont {Chen}}, \bibinfo {author} {\bibfnamefont {M.}~\bibnamefont
  {Li}}, \bibinfo {author} {\bibfnamefont {G.}~\bibnamefont {Wu}},\ and\
  \bibinfo {author} {\bibfnamefont {B.}~\bibnamefont {Luo}},\ }\href
  {https://doi.org/10.1063/5.0194784} {\bibfield  {journal} {\bibinfo
  {journal} {Appl. Phys. Lett.}\ }\textbf {\bibinfo {volume} {124}},\ \bibinfo
  {pages} {084002} (\bibinfo {year} {2024})}\BibitemShut {NoStop}%
\bibitem [{\citenamefont {Happer}(1972)}]{happer1972optical}%
  \BibitemOpen
  \bibfield  {author} {\bibinfo {author} {\bibfnamefont {W.}~\bibnamefont
  {Happer}},\ }\href {https://doi.org/10.1103/RevModPhys.44.169} {\bibfield
  {journal} {\bibinfo  {journal} {Rev. Mod. Phys.}\ }\textbf {\bibinfo {volume}
  {44}},\ \bibinfo {pages} {169} (\bibinfo {year} {1972})}\BibitemShut
  {NoStop}%
\bibitem [{\citenamefont {Berman}\ \emph {et~al.}(1982)\citenamefont {Berman},
  \citenamefont {Mossberg},\ and\ \citenamefont
  {Hartmann}}]{berman1982collision}%
  \BibitemOpen
  \bibfield  {author} {\bibinfo {author} {\bibfnamefont {P.~R.}\ \bibnamefont
  {Berman}}, \bibinfo {author} {\bibfnamefont {T.~W.}\ \bibnamefont
  {Mossberg}},\ and\ \bibinfo {author} {\bibfnamefont {S.~R.}\ \bibnamefont
  {Hartmann}},\ }\href {https://doi.org/10.1103/PhysRevA.25.2550} {\bibfield
  {journal} {\bibinfo  {journal} {Phys. Rev. A}\ }\textbf {\bibinfo {volume}
  {25}},\ \bibinfo {pages} {2550} (\bibinfo {year} {1982})}\BibitemShut
  {NoStop}%
\bibitem [{\citenamefont {Firstenberg}\ \emph {et~al.}(2013)\citenamefont
  {Firstenberg}, \citenamefont {Shuker}, \citenamefont {Ron},\ and\
  \citenamefont {Davidson}}]{firstenberg2013coherent}%
  \BibitemOpen
  \bibfield  {author} {\bibinfo {author} {\bibfnamefont {O.}~\bibnamefont
  {Firstenberg}}, \bibinfo {author} {\bibfnamefont {M.}~\bibnamefont {Shuker}},
  \bibinfo {author} {\bibfnamefont {A.}~\bibnamefont {Ron}},\ and\ \bibinfo
  {author} {\bibfnamefont {N.}~\bibnamefont {Davidson}},\ }\href
  {https://doi.org/10.1103/RevModPhys.85.941} {\bibfield  {journal} {\bibinfo
  {journal} {Rev. Mod. Phys.}\ }\textbf {\bibinfo {volume} {85}},\ \bibinfo
  {pages} {941} (\bibinfo {year} {2013})}\BibitemShut {NoStop}%
\bibitem [{\citenamefont {Maguire}\ \emph {et~al.}(2006)\citenamefont
  {Maguire}, \citenamefont {van Bijnen}, \citenamefont {Mese},\ and\
  \citenamefont {Scholten}}]{maguire2006theoretical}%
  \BibitemOpen
  \bibfield  {author} {\bibinfo {author} {\bibfnamefont {L.~P.}\ \bibnamefont
  {Maguire}}, \bibinfo {author} {\bibfnamefont {R.~M.~W.}\ \bibnamefont {van
  Bijnen}}, \bibinfo {author} {\bibfnamefont {E.}~\bibnamefont {Mese}},\ and\
  \bibinfo {author} {\bibfnamefont {R.~E.}\ \bibnamefont {Scholten}},\ }\href
  {https://doi.org/10.1088/0953-4075/39/12/007} {\bibfield  {journal} {\bibinfo
   {journal} {J. Phys. B: At. Mol. Opt. Phys.}\ }\textbf {\bibinfo {volume}
  {39}},\ \bibinfo {pages} {2709} (\bibinfo {year} {2006})}\BibitemShut
  {NoStop}%
\bibitem [{\citenamefont {Lindvall}\ and\ \citenamefont
  {Tittonen}(2009)}]{lindvall2009interaction}%
  \BibitemOpen
  \bibfield  {author} {\bibinfo {author} {\bibfnamefont {T.}~\bibnamefont
  {Lindvall}}\ and\ \bibinfo {author} {\bibfnamefont {I.}~\bibnamefont
  {Tittonen}},\ }\href {https://doi.org/10.1103/PhysRevA.80.032505} {\bibfield
  {journal} {\bibinfo  {journal} {Phys. Rev. A}\ }\textbf {\bibinfo {volume}
  {80}},\ \bibinfo {pages} {032505} (\bibinfo {year} {2009})}\BibitemShut
  {NoStop}%
\bibitem [{\citenamefont {van Lange}\ \emph {et~al.}(2020)\citenamefont {van
  Lange}, \citenamefont {van~der Straten},\ and\ \citenamefont {van
  Oosten}}]{vanlange2020combined}%
  \BibitemOpen
  \bibfield  {author} {\bibinfo {author} {\bibfnamefont {A.~J.}\ \bibnamefont
  {van Lange}}, \bibinfo {author} {\bibfnamefont {P.}~\bibnamefont {van~der
  Straten}},\ and\ \bibinfo {author} {\bibfnamefont {D.}~\bibnamefont {van
  Oosten}},\ }\href {https://doi.org/10.1088/1361-6455/ab7fc2} {\bibfield
  {journal} {\bibinfo  {journal} {J. Phys. B: At. Mol. Opt. Phys.}\ }\textbf
  {\bibinfo {volume} {53}},\ \bibinfo {pages} {125402} (\bibinfo {year}
  {2020})}\BibitemShut {NoStop}%
\bibitem [{\citenamefont {H{\"a}upl}\ \emph {et~al.}(2025)\citenamefont
  {H{\"a}upl}, \citenamefont {Higgins}, \citenamefont {Pizzey}, \citenamefont
  {Briscoe}, \citenamefont {Wrathmall}, \citenamefont {Hughes}, \citenamefont
  {L{\"o}w},\ and\ \citenamefont {Joly}}]{haupl2025modelling}%
  \BibitemOpen
  \bibfield  {author} {\bibinfo {author} {\bibfnamefont {D.~R.}\ \bibnamefont
  {H{\"a}upl}}, \bibinfo {author} {\bibfnamefont {C.~R.}\ \bibnamefont
  {Higgins}}, \bibinfo {author} {\bibfnamefont {D.}~\bibnamefont {Pizzey}},
  \bibinfo {author} {\bibfnamefont {J.~D.}\ \bibnamefont {Briscoe}}, \bibinfo
  {author} {\bibfnamefont {S.~A.}\ \bibnamefont {Wrathmall}}, \bibinfo {author}
  {\bibfnamefont {I.~G.}\ \bibnamefont {Hughes}}, \bibinfo {author}
  {\bibfnamefont {R.}~\bibnamefont {L{\"o}w}},\ and\ \bibinfo {author}
  {\bibfnamefont {N.~Y.}\ \bibnamefont {Joly}},\ }\href
  {https://doi.org/10.1088/1367-2630/adb77c} {\bibfield  {journal} {\bibinfo
  {journal} {New J. Phys.}\ }\textbf {\bibinfo {volume} {27}},\ \bibinfo
  {pages} {033003} (\bibinfo {year} {2025})}\BibitemShut {NoStop}%
\bibitem [{\citenamefont {Torrey}(1956)}]{torrey1956bloch}%
  \BibitemOpen
  \bibfield  {author} {\bibinfo {author} {\bibfnamefont {H.~C.}\ \bibnamefont
  {Torrey}},\ }\href {https://doi.org/10.1103/PhysRev.104.563} {\bibfield
  {journal} {\bibinfo  {journal} {Phys. Rev.}\ }\textbf {\bibinfo {volume}
  {104}},\ \bibinfo {pages} {563} (\bibinfo {year} {1956})}\BibitemShut
  {NoStop}%
\bibitem [{\citenamefont {Firstenberg}\ \emph {et~al.}(2008)\citenamefont
  {Firstenberg}, \citenamefont {Shuker}, \citenamefont {Pugatch}, \citenamefont
  {Fredkin}, \citenamefont {Davidson},\ and\ \citenamefont
  {Ron}}]{firstenberg2008theory}%
  \BibitemOpen
  \bibfield  {author} {\bibinfo {author} {\bibfnamefont {O.}~\bibnamefont
  {Firstenberg}}, \bibinfo {author} {\bibfnamefont {M.}~\bibnamefont {Shuker}},
  \bibinfo {author} {\bibfnamefont {R.}~\bibnamefont {Pugatch}}, \bibinfo
  {author} {\bibfnamefont {D.~R.}\ \bibnamefont {Fredkin}}, \bibinfo {author}
  {\bibfnamefont {N.}~\bibnamefont {Davidson}},\ and\ \bibinfo {author}
  {\bibfnamefont {A.}~\bibnamefont {Ron}},\ }\href
  {https://doi.org/10.1103/PhysRevA.77.043830} {\bibfield  {journal} {\bibinfo
  {journal} {Phys. Rev. A}\ }\textbf {\bibinfo {volume} {77}},\ \bibinfo
  {pages} {043830} (\bibinfo {year} {2008})}\BibitemShut {NoStop}%
\bibitem [{\citenamefont {Dicke}(1953)}]{dicke1953collisions}%
  \BibitemOpen
  \bibfield  {author} {\bibinfo {author} {\bibfnamefont {R.~H.}\ \bibnamefont
  {Dicke}},\ }\href {https://doi.org/10.1103/PhysRev.89.472} {\bibfield
  {journal} {\bibinfo  {journal} {Phys. Rev.}\ }\textbf {\bibinfo {volume}
  {89}},\ \bibinfo {pages} {472} (\bibinfo {year} {1953})}\BibitemShut
  {NoStop}%
\bibitem [{\citenamefont {Firstenberg}\ \emph {et~al.}(2007)\citenamefont
  {Firstenberg}, \citenamefont {Shuker}, \citenamefont {Ben-Kish},
  \citenamefont {Fredkin}, \citenamefont {Davidson},\ and\ \citenamefont
  {Ron}}]{firstenberg2007dicke}%
  \BibitemOpen
  \bibfield  {author} {\bibinfo {author} {\bibfnamefont {O.}~\bibnamefont
  {Firstenberg}}, \bibinfo {author} {\bibfnamefont {M.}~\bibnamefont {Shuker}},
  \bibinfo {author} {\bibfnamefont {A.}~\bibnamefont {Ben-Kish}}, \bibinfo
  {author} {\bibfnamefont {D.~R.}\ \bibnamefont {Fredkin}}, \bibinfo {author}
  {\bibfnamefont {N.}~\bibnamefont {Davidson}},\ and\ \bibinfo {author}
  {\bibfnamefont {A.}~\bibnamefont {Ron}},\ }\href
  {https://doi.org/10.1103/PhysRevA.76.013818} {\bibfield  {journal} {\bibinfo
  {journal} {Phys. Rev. A}\ }\textbf {\bibinfo {volume} {76}},\ \bibinfo
  {pages} {013818} (\bibinfo {year} {2007})}\BibitemShut {NoStop}%
\bibitem [{\citenamefont {Bord\'{e}}\ \emph {et~al.}(1976)\citenamefont
  {Bord\'{e}}, \citenamefont {Hall}, \citenamefont {Kunasz},\ and\
  \citenamefont {Hummer}}]{borde1976transit}%
  \BibitemOpen
  \bibfield  {author} {\bibinfo {author} {\bibfnamefont {C.~J.}\ \bibnamefont
  {Bord\'{e}}}, \bibinfo {author} {\bibfnamefont {J.~L.}\ \bibnamefont {Hall}},
  \bibinfo {author} {\bibfnamefont {C.~V.}\ \bibnamefont {Kunasz}},\ and\
  \bibinfo {author} {\bibfnamefont {D.~G.}\ \bibnamefont {Hummer}},\ }\href
  {https://doi.org/10.1103/PhysRevA.14.236} {\bibfield  {journal} {\bibinfo
  {journal} {Phys. Rev. A}\ }\textbf {\bibinfo {volume} {14}},\ \bibinfo
  {pages} {236} (\bibinfo {year} {1976})}\BibitemShut {NoStop}%
\bibitem [{\citenamefont {Chekhovskoi}\ and\ \citenamefont
  {Labzowsky}(2024)}]{chekhovskoi2024transit}%
  \BibitemOpen
  \bibfield  {author} {\bibinfo {author} {\bibfnamefont {S.~D.}\ \bibnamefont
  {Chekhovskoi}}\ and\ \bibinfo {author} {\bibfnamefont {L.~N.}\ \bibnamefont
  {Labzowsky}},\ }\href {https://doi.org/10.1103/PhysRevA.109.062811}
  {\bibfield  {journal} {\bibinfo  {journal} {Phys. Rev. A}\ }\textbf {\bibinfo
  {volume} {109}},\ \bibinfo {pages} {062811} (\bibinfo {year}
  {2024})}\BibitemShut {NoStop}%
\bibitem [{\citenamefont {Xiao}\ \emph {et~al.}(2006)\citenamefont {Xiao},
  \citenamefont {Novikova}, \citenamefont {Phillips},\ and\ \citenamefont
  {Walsworth}}]{xiao2006diffusion}%
  \BibitemOpen
  \bibfield  {author} {\bibinfo {author} {\bibfnamefont {Y.}~\bibnamefont
  {Xiao}}, \bibinfo {author} {\bibfnamefont {I.}~\bibnamefont {Novikova}},
  \bibinfo {author} {\bibfnamefont {D.~F.}\ \bibnamefont {Phillips}},\ and\
  \bibinfo {author} {\bibfnamefont {R.~L.}\ \bibnamefont {Walsworth}},\ }\href
  {https://doi.org/10.1103/PhysRevLett.96.043601} {\bibfield  {journal}
  {\bibinfo  {journal} {Phys. Rev. Lett.}\ }\textbf {\bibinfo {volume} {96}},\
  \bibinfo {pages} {043601} (\bibinfo {year} {2006})}\BibitemShut {NoStop}%
\bibitem [{\citenamefont {Tilchin}\ \emph {et~al.}(2011)\citenamefont
  {Tilchin}, \citenamefont {Wilson-Gordon},\ and\ \citenamefont
  {Firstenberg}}]{tilchin2011thermal}%
  \BibitemOpen
  \bibfield  {author} {\bibinfo {author} {\bibfnamefont {E.}~\bibnamefont
  {Tilchin}}, \bibinfo {author} {\bibfnamefont {A.~D.}\ \bibnamefont
  {Wilson-Gordon}},\ and\ \bibinfo {author} {\bibfnamefont {O.}~\bibnamefont
  {Firstenberg}},\ }\href {https://doi.org/10.1103/PhysRevA.83.053812}
  {\bibfield  {journal} {\bibinfo  {journal} {Phys. Rev. A}\ }\textbf {\bibinfo
  {volume} {83}},\ \bibinfo {pages} {053812} (\bibinfo {year}
  {2011})}\BibitemShut {NoStop}%
\bibitem [{\citenamefont {Shuker}\ \emph {et~al.}(2008)\citenamefont {Shuker},
  \citenamefont {Firstenberg}, \citenamefont {Pugatch}, \citenamefont {Ron},\
  and\ \citenamefont {Davidson}}]{shuker2008storing}%
  \BibitemOpen
  \bibfield  {author} {\bibinfo {author} {\bibfnamefont {M.}~\bibnamefont
  {Shuker}}, \bibinfo {author} {\bibfnamefont {O.}~\bibnamefont {Firstenberg}},
  \bibinfo {author} {\bibfnamefont {R.}~\bibnamefont {Pugatch}}, \bibinfo
  {author} {\bibfnamefont {A.}~\bibnamefont {Ron}},\ and\ \bibinfo {author}
  {\bibfnamefont {N.}~\bibnamefont {Davidson}},\ }\href
  {https://doi.org/10.1103/PhysRevLett.100.223601} {\bibfield  {journal}
  {\bibinfo  {journal} {Phys. Rev. Lett.}\ }\textbf {\bibinfo {volume} {100}},\
  \bibinfo {pages} {223601} (\bibinfo {year} {2008})}\BibitemShut {NoStop}%
\bibitem [{\citenamefont {Firstenberg}\ \emph {et~al.}(2010)\citenamefont
  {Firstenberg}, \citenamefont {London}, \citenamefont {Yankelev},
  \citenamefont {Pugatch}, \citenamefont {Shuker},\ and\ \citenamefont
  {Davidson}}]{firstenberg2010selfsimilar}%
  \BibitemOpen
  \bibfield  {author} {\bibinfo {author} {\bibfnamefont {O.}~\bibnamefont
  {Firstenberg}}, \bibinfo {author} {\bibfnamefont {P.}~\bibnamefont {London}},
  \bibinfo {author} {\bibfnamefont {D.}~\bibnamefont {Yankelev}}, \bibinfo
  {author} {\bibfnamefont {R.}~\bibnamefont {Pugatch}}, \bibinfo {author}
  {\bibfnamefont {M.}~\bibnamefont {Shuker}},\ and\ \bibinfo {author}
  {\bibfnamefont {N.}~\bibnamefont {Davidson}},\ }\href
  {https://doi.org/10.1103/PhysRevLett.105.183602} {\bibfield  {journal}
  {\bibinfo  {journal} {Phys. Rev. Lett.}\ }\textbf {\bibinfo {volume} {105}},\
  \bibinfo {pages} {183602} (\bibinfo {year} {2010})}\BibitemShut {NoStop}%
\bibitem [{\citenamefont {Finkelstein}\ \emph {et~al.}(2023)\citenamefont
  {Finkelstein}, \citenamefont {Bali}, \citenamefont {Firstenberg},\ and\
  \citenamefont {Novikova}}]{finkelstein2023practical}%
  \BibitemOpen
  \bibfield  {author} {\bibinfo {author} {\bibfnamefont {R.}~\bibnamefont
  {Finkelstein}}, \bibinfo {author} {\bibfnamefont {S.}~\bibnamefont {Bali}},
  \bibinfo {author} {\bibfnamefont {O.}~\bibnamefont {Firstenberg}},\ and\
  \bibinfo {author} {\bibfnamefont {I.}~\bibnamefont {Novikova}},\ }\href
  {https://doi.org/10.1088/1367-2630/acbc40} {\bibfield  {journal} {\bibinfo
  {journal} {New J. Phys.}\ }\textbf {\bibinfo {volume} {25}},\ \bibinfo
  {pages} {035001} (\bibinfo {year} {2023})}\BibitemShut {NoStop}%
\bibitem [{\citenamefont {Steck}(2025)}]{steck2025rb85}%
  \BibitemOpen
  \bibfield  {author} {\bibinfo {author} {\bibfnamefont {D.~A.}\ \bibnamefont
  {Steck}},\ }\href {https://steck.us/alkalidata} {\bibinfo {title} {Rubidium
  85 d line data}},\ \bibinfo {howpublished} {Available online at
  \url{https://steck.us/alkalidata}} (\bibinfo {year} {2025}),\ \bibinfo {note}
  {revision 2.3.4, 8 August 2025}\BibitemShut {NoStop}%
\bibitem [{\citenamefont {Yeh}(1982)}]{yeh1982dispersive}%
  \BibitemOpen
  \bibfield  {author} {\bibinfo {author} {\bibfnamefont {P.}~\bibnamefont
  {Yeh}},\ }\href {https://doi.org/10.1364/AO.21.002069} {\bibfield  {journal}
  {\bibinfo  {journal} {Appl. Opt.}\ }\textbf {\bibinfo {volume} {21}},\
  \bibinfo {pages} {2069} (\bibinfo {year} {1982})}\BibitemShut {NoStop}%
\bibitem [{\citenamefont {Yin}\ and\ \citenamefont
  {Shay}(1991)}]{yin1991theoretical}%
  \BibitemOpen
  \bibfield  {author} {\bibinfo {author} {\bibfnamefont {B.}~\bibnamefont
  {Yin}}\ and\ \bibinfo {author} {\bibfnamefont {T.~M.}\ \bibnamefont {Shay}},\
  }\href {https://doi.org/10.1364/OL.16.001617} {\bibfield  {journal} {\bibinfo
   {journal} {Opt. Lett.}\ }\textbf {\bibinfo {volume} {16}},\ \bibinfo {pages}
  {1617} (\bibinfo {year} {1991})}\BibitemShut {NoStop}%
\bibitem [{\citenamefont {Olsen}\ \emph {et~al.}(2011)\citenamefont {Olsen},
  \citenamefont {Patton}, \citenamefont {Jau},\ and\ \citenamefont
  {Happer}}]{olsen2011optical}%
  \BibitemOpen
  \bibfield  {author} {\bibinfo {author} {\bibfnamefont {B.~A.}\ \bibnamefont
  {Olsen}}, \bibinfo {author} {\bibfnamefont {B.}~\bibnamefont {Patton}},
  \bibinfo {author} {\bibfnamefont {Y.-Y.}\ \bibnamefont {Jau}},\ and\ \bibinfo
  {author} {\bibfnamefont {W.}~\bibnamefont {Happer}},\ }\href
  {https://doi.org/10.1103/PhysRevA.84.063410} {\bibfield  {journal} {\bibinfo
  {journal} {Phys. Rev. A}\ }\textbf {\bibinfo {volume} {84}},\ \bibinfo
  {pages} {063410} (\bibinfo {year} {2011})}\BibitemShut {NoStop}%
\end{thebibliography}%

\end{document}